\documentclass[11pt]{article}

\usepackage{tabularx} 
\usepackage{amsmath}  
\usepackage{graphicx} 
\usepackage{amsmath}
\usepackage{floatrow}
\usepackage{placeins}
\usepackage[utf8]{inputenc}
\usepackage[T1]{fontenc}
\usepackage{authblk}
\usepackage{soul}
\usepackage{dblfloatfix}

\usepackage[justification=centering]{caption}
\usepackage{caption}
\usepackage{color}

\usepackage[margin=1in,letterpaper]{geometry} 
\usepackage[final]{hyperref} 
\hypersetup{
	colorlinks=true,       
	linkcolor=blue,        
	citecolor=blue,        
	filecolor=magenta,     
	urlcolor=blue         
}

\begin{document}
\newcommand\tab[1][0.5cm]{\hspace*{#1}}

\title{Computer simulation of surgical interventions \\for the treatment of refractory pulmonary hypertension}
\author[1,3]{Seong Woo Han}
\author[1,2]{Charles Puelz}
\author[2]{Craig G. Rusin}
\author[2]{\\Daniel J. Penny}
\author[4]{Ryan Coleman}
\author[1]{Charles S. Peskin}
\affil[1]{Courant Institute of Mathematical Sciences, New York University}
\affil[2]{Department of Pediatrics, Section of Cardiology, Baylor College of Medicine and Texas Children's Hospital}
\affil[3]{Department of Bioengineering, University of Pennsylvania}
\affil[4]{Department of Pediatrics, Section of Critical Care Medicine, Baylor College of Medicine and Texas Children's Hospital}


\maketitle

\begin{abstract}

This paper describes computer models of three interventions used for treating refractory pulmonary hypertension (RPH). These procedures create either an atrial septal defect, a ventricular septal defect, or, in the case of a Potts shunt, a patent ductus arteriosus. The aim in all three cases is to generate a right-to-left shunt, allowing for either pressure or volume unloading of the right side of the heart in the setting of right ventricular failure, while maintaining cardiac output. These shunts are created, however, at the expense of introducing de-oxygenated blood into the systemic circulation, thereby lowering the systemic arterial oxygen saturation. The models developed in this paper are based on compartmental descriptions of human hemodynamics and oxygen transport. An important parameter included in our models is the cross-sectional area of the surgically created defect. Numerical simulations are performed to compare different interventions and various shunt sizes and to assess their impact on hemodynamic variables and oxygen saturations. We also create a model for exercise and use it to study exercise tolerance in simulated pre-intervention and post-intervention RPH patients.
\end{abstract}

\section{Introduction}
Pulmonary hypertension refers to a spectrum of cardiovascular and/or pulmonary diseases that involve elevations in a person’s  pulmonary vascular resistance (PVR).  Over time, this elevated PVR causes pathologic remodeling of the right ventricle and the pulmonary vasculature, ultimately resulting in right ventricular failure and death. In this paper, our focus is on interventions for refractory pulmonary hypertension (RPH), which corresponds to disease that is unresponsive to standard medical treatments \cite{Etz07, Levy11}. Even with aggressive pharmacotherapy, patients often will ultimately require either lung transplantation or palliative surgical or catheter-based procedures, with the goal of either approach being to provide relief to the ailing right ventricle and thereby to extend the patient's life, although for an unknown period of time. 


Palliative shunts used in the setting of a failing right ventricle due to elevated PVR can be classified into two main categories: (1) pre-tricuspid shunts, meaning the shunt occurs prior to blood crossing the tricuspid valve, which is what occurs in the setting of an atrial septal defect (ASD), and (2) post-tricuspid shunts, meaning the shunt occurs after the blood passes the tricuspid valve, which are where the ventricular septal defect (VSD) and Potts shunt occur.  This classification scheme is important, as pre-tricuspid shunts are viewed as volume-unloading shunts for the right ventricle, meaning they can unload excess volume from a failing right ventricle, but do not directly affect the pressure the right ventricle has to pump against.  Post-tricuspid shunts, on the other hand, are pressure-unloading shunts, meaning they provide a lower-resistance pathway for blood to traverse, decreasing the resistance the failing right ventricle has to pump against.  Pressure-unloading shunts are often preferred, as it is the pressure load the right ventricle has to pump against that results in its failure and the patient's demise.  Because these shunts, regardless of location, allow for right-to-left shunting, they result in a decrease in systemic oxygen saturations of varying degrees and severity. Shunt effectiveness is determined not only by location of the shunt, but by the size of the shunt as well.  Small septal defects may quickly become restrictive over time, reducing their effectiveness at either pressure- or volume-unloading the right ventricle.  The same is true for narrow shunts like a Potts, particularly as their length increases.  


The effects of various shunts on pressures, flows, and oxygen saturations are often not clear in practice. Furthermore, shunt flows are highly sensitive to shunt size, a parameter that can be varied within the modeling framework developed in this paper.  The goal of this paper is to use computational models to study the impact of several possible shunts, used for treating refractory pulmonary hypertension, on important hemodynamic variables and oxygen saturations. The three shunts considered here are (1) within the atrial septum, (2) within the ventricular septum, or (3) between the pulmonary artery and the aorta \cite{Leopold15, Baruteau15}. We refer to these surgically created defects respectively by the following names: (1) an atrial septal defect (ASD), (2) a ventricular septal defect (VSD), or (3) a Potts shunt. For each intervention, we develop and apply computational models to study both the benefit in terms of reduced pulmonary artery pressure and also the detriment in terms of systemic arterial oxygen desaturation, as functions of the cross-sectional area of the shunt. Furthermore, we develop an exercise model and use it to study the simulated impact of RPH on exercise tolerance in three distinct conditions: pre-intervention, immediately post-intervention, and a certain amount of time post-intervention after which pulmonary vascular remodeling has occurred.


The complexities associated with pulmonary hypertension have motivated the use of computational models for studying disease progression, diagnosis, and the performance of possible treatments. We recall several important contributions of physics-based models for pulmonary hypertension. Delhass et al. constructed compartmental models for two possible interventions considered in this paper, the ASD and the Potts shunt \cite{Delhaas18}.  Our paper extends their results to a comparison of the ASD and Potts shunt with the VSD.  Gerringer et al. built and calibrated compartmental models  with animal data to study the effect of progressing pulmonary hypertension on resistance and compliance \cite{Gerringer18}.  Tewari et al. also used calibrated compartmental models to investigate changes in important hemodynamic parameters after the onset of disease \cite{Tewari13}.  Vessel network models, which incorporate spatial variations of blood flow and pressure, were used by Acosta et al. to derive early diagnostic indicators of disease \cite{Acosta17}.  Qureshi et al. also used vessel network models to study several classes of pulmonary hypertension as well as to simulate control and diseased animal models \cite{Qureshi14, Qureshi19}. There have also been modeling efforts to understand the impact of pulmonary hypertension on remodeling of heart tissue.  Raush et al. constructed three-dimensional solid mechanics models of the ventricular chambers that were coupled to a mathematical description of tissue remodeling under the high pressure loads associated with hypertension \cite{Rausch11}.

The rest of this paper is organized as follows.  Section \ref{sec:models} describes the mathematical models for blood flow and oxygen transport and the numerical methods used to approximate the resulting equations.  This section also includes a discussion of parameter selection, the shunt model derived from the Gorlin equation, and the exercise model.  Section \ref{sec:results} describes results from our models for each of the three possible surgical interventions. Our models are used to study the dependence of several important hemodynamic variables on shunt size. We also investigate mean shunt flow as a function of pulmonary vascular resistance and the corresponding time-dependent details of the shunt flow waveform.  Limitations and conclusions are provided in Sections \ref{sec:limitations} and \ref{sec:conc}.

\section{Circulation models and numerical methods}
\label{sec:models}

In this section, we present the models used in this work and the numerical methods by which the model equations are solved. The following subsections describe a hemodynamic model, a cardiac chamber model that specifies the time-varying compliance of each heart chamber in the hemodynamic model, a shunt model based on the Gorlin equation that makes it possible to include shunts of specified cross-sectional area, an oxygen transport model that calculates oxygen saturations throughout the circulation, and finally an exercise model that modulates several hemodynamic parameters to simulate the impact of exercise on blood flow and oxygen transport.

\subsection{Hemodynamic model}
\label{subsec:hemo}
The circulation is represented by a collection of compartments corresponding to compliance chambers.  These chambers are connected by resistors that are equipped with valves \cite{Peskin86}. We use the following compliance relation for each of the $N$ compliance chambers, numbered $i = 1,2,...,N$:
\begin{equation}
    V_i = (V_\text{d})_i + C_i P_i, \quad i = 1,...,N.
    \label{eq:compliance}
\end{equation}
The parameter $C_{i}$ is the compliance of chamber $i$, which is assumed to be constant for arteries and veins but time-varying for the heart chambers. The variable $V_{i}$ is the volume of compliance chamber $i$, the variable $P_{i}$ is the pressure of that chamber, and the parameter $(V_\text{d})_{i}$ is the dead volume, that is, the volume of the chamber when the pressure is zero. We assume the flow from chamber $i$ to chamber $j$ is governed by a pressure-flow relationship of the following form:
\begin{equation}
    Q_{ij} = \frac{S_{ij}}{R_{ij}}(P_{i}-P_{j}) = S_{ij}\,G_{ij}\,(P_{i}-P_{j}), \quad i,j = 1,...,N,
    \label{eq:pressflow}
\end{equation}
where 
\begin{equation}
S_{ij} = 
 \begin{cases} 
      1, & P_i > P_j, \\
      0, & P_i\leq P_j. 
 \end{cases}
 \label{eq:valvestate}
\end{equation}\\
Equation \eqref{eq:pressflow} describes the flow through a resistance that is equipped with a valve. The conductance $G_{ij}$ is the reciprocal of the resistance $R_{ij}$. Conductance is convenient because it can be set equal to zero to represent the absence of a connection between two chambers. The variable $S_{ij}$, which is determined by $P_{i}$ and $P_{j}$ according to equation \eqref{eq:valvestate}, denotes the state of the valve, with $S_{ij}$ = 1 when the valve is open, and $S_{ij}$ = 0 when the valve is closed. Note that the words ``open'' and ``closed'' have the opposite meaning here from their use in electricity, where a closed switch is conducting and an open switch is non-conducting.

Equipping \textit{every} connection with a valve does not involve any loss of generality. Between any pair of chambers $i$ and $j$, our framework allows for \textit{two} connections of the type described above, one with a valve that allows flow only from $i$ to $j$, and another with a valve that allows flow only from $j$ to $i$. To model a situation in which there is no valve in a connection between chambers $i$ and $j$, we need only set $G_{ij}$ equal to $G_{ji}$. To model a leaky valve we may set $G_{ij}$ and $G_{ji}$ to positive but unequal values. Lastly, to model the situation in which there is no connection at all between chambers $i$ and $j$, we set $G_{ij}$ = $G_{ji}$ = 0. Thus, our framework allows for a great variety of connection types and patterns merely by specifying the (non-symmetric) $N$ by $N$ matrix $G$.

Upon differentiating equation \eqref{eq:compliance} with respect to time and using the principle that the rate of change of volume is equal to inflow minus outflow, together with equation \eqref{eq:pressflow}, one obtains the following system of ordinary differential equations for the pressures as functions of time:
\begin{align}
    \frac{d}{dt}(C_{i}P_{i}) & = \displaystyle\sum_{j=1}^{N} (S_{ji}G_{ji}(P_{j} - P_{i}) - S_{ij}G_{ij}(P_{i} - P_{j})) \nonumber\\
    & = \displaystyle\sum_{j=1}^{N} (S_{ij}G_{ij}+S_{ji}G_{ji})(P_{j} - P_{i}).
    \label{eq:diffeq}
\end{align}
We assume here that all of the dead volumes are constant but allow for the possibility that some of the compliances, specifically those of the heart chambers, are functions of time. How these compliances are specified will be described in Subsection \ref{subsec:chamber}. Equation \eqref{eq:diffeq} will be modified later to include shunt flows modeled by the Gorlin equation; see Subsection \ref{subsec:shunt}.

Our numerical scheme for equation \eqref{eq:diffeq} is the backward Euler method:
\begin{align}
    \frac{C_{i}^nP_{i}^n - C_{i}^{n-1}P_{i}^{n-1}}{\Delta t} & = \displaystyle\sum_{j=1}^{N} (S_{ij}^nG_{ij}^n + S_{ji}^{n}G_{ji}^n)(P_{j}^n - P_{i}^n).
    \label{eq:backwardeuler}
\end{align}
This is a system of equations for the unknown pressures at time step $n$. It is a nonlinear system, because $S_{ij}$ is a function of $P_{i}$ and $P_{j}$, see equation \eqref{eq:valvestate}. The reason for using the backward Euler method here is its unconditional stability. If two compliance chambers are connected by a very large conductance (that is, by a very small resistance), their pressures will equilibrate on a very fast time scale, and we do not want to be required to use a small enough time step to resolve the details of that rapid equilibration. This situation actually arises in the circulation whenever there are two chambers with an open heart valve between them, since an open valve (at least when it is non-stenotic) has a very high conductance. 

The procedure that we use to solve the nonlinear system \eqref{eq:backwardeuler} is based on the following observation: given the valve states, equation \eqref{eq:backwardeuler} reduces to a linear system that is easy to solve for the pressures. Also, given the pressures, it is easy to evaluate the valve states from equation \eqref{eq:valvestate}. Thus, the procedure starts with a guess for the valve states (a good guess is the valve states that were found on the previous time step), solves equation \eqref{eq:backwardeuler} for the pressures, resets the valve states according to the pressures via equation \eqref{eq:valvestate}, and so on. The process stops when the valve states (and therefore the pressures) stop changing, and in practice this happens very quickly. On most time steps, the initial guess, that the valves states are the same as they were at the previous time step, turns out to be correct. When the valve states stop changing, the problem stated in equation \eqref{eq:backwardeuler} is actually solved (except, of course, for round-off error), not merely solved to within some tolerance. This is because the valve states are discrete.

For further discussion of the methodology described here, see \cite{Hoppensteadt12}. As in that reference, our models use six compliance chambers corresponding to the left and right ventricles and the systemic and pulmonary arteries and veins. We do not separately model the atria, but instead treat each atrium as part of the venous system to which it is connected, and moreover we do not take into account the time dependence of the atrial compliances. The ventricular compliances are, of course, time dependent in our model, but not in the same way as in \cite{Hoppensteadt12}, see Subsection \ref{subsec:chamber}.

\begin{table}
\begin{center}
\label{TestTable}
\begin{tabular}{l*{6}{c}r}
\hline
Parameters               & Resistance ($R$) & Dead Volume ($V_\text{d}$) & Compliance ($C$) \\
\hline
Units     & mmHg/(L/min) & L & L/mmHg  \\
\hline
S           & 17.5 & - & -    \\
P           & 1.79, 22.75 & - & -   \\
Mi           & 0.01 & - & -  \\
Ao           & 0.01 & - & -     \\
Tr           & 0.01 & - & -    \\
Pu           & 0.01 & - & -    \\
SA          & - & 0.825 &  0.0012    \\
PA         & - & 0.1135 & 0.0042, 0.0021    \\
PV        & - & 0.18 & 0.01   \\
SV        & - & 3.1 & 0.09   \\


\hline
\end{tabular}
\caption{Parameters for the circulation models. When two numbers are given, the first one is used in the normal model and the second one is used in the RPH model. For example, pulmonary resistance is chosen to be 1.3 times greater than the systemic resistance (1.3$\times$17.5) to simulate refractory pulmonary hypertension.  The parameters shown in the table are resting values. The systemic venous dead volume and systemic resistance are modified by our exercise model as discussed in section \ref{sec:exercise_models}. Abbreviations: S, systemic organs; P, lungs; Mi, mitral valve; Ao, aortic valve; Tr, tricuspid valve; Pu, pulmonic valve; SA, systemic arteries; PA, pulmonary arteries; PV, pulmonary veins; SV, systemic veins.}
\label{table:parms1}
\end{center}
\end{table}

To create a model for RPH, we first determine parameter values that result in a normal model for a healthy circulation. Then, parameters for the normal model are systematically adjusted, as described below, to create a pre-intervention model corresponding to the RPH disease state. In order to model a severe pulmonary hypertension, the pulmonary resistance is taken to be 1.3 times greater than the systemic resistance \cite{Roy13}. This is very different from the normal case in which the pulmonary resistance is approximately 10 times \textit{smaller} than the systemic resistance \cite{Kovacs12, Naderi18, Widrich20}. A possible consequence of pulmonary hypertension is right-heart hypertrophy, making the normally thin-walled right ventricle into a chamber that more closely resembles the normal left ventricle \cite{Ryan14, Noordegraaf19, Chemla02}. To model this, we use typical left-ventricular parameters for both ventricles; see the next subsection. Another consequence of pulmonary hypertension is right heart failure that leads to increased blood volume \cite{Stickel19}. Accordingly, we use a total blood volume of 5.248 L, compared to the blood volume in our normal model, which is 5.098 L. This change is needed to elevate the systemic venous pressure sufficiently to fill the hypertrophied right heart and produce a viable cardiac output. The hemodynamic parameters used in the present model, other than the cardiac chamber parameters, are stated in Table \ref{table:parms1}. When two values are given, the first corresponds to the normal model and the second corresponds to the pre-intervention RPH model. 

\subsection{Cardiac chamber model}
\label{subsec:chamber}

This subsection details the time-varying elastance model used for the left and right ventricles, adapted from \cite{Mynard12}. Note that the elastance, denoted by $E$, is the reciprocal of the compliance $C$. For a cardiac chamber, the elastance, and therefore the compliance, is a given function of time. For the pre-intervention RPH model, we use the same elastance function $E(t)$, with the same parameters, for the left and right ventricles.  This choice is reasonable because the large right-sided pressures associated with pulmonary hypertension lead to remodeling and thickening of the right ventricular wall \cite{Giusca16}.  In severe pulmonary hypertension, these changes result in a right ventricular pressure/volume characteristics similar to that of the left ventricle \cite{Ryo15}. Maximum and minimum ventricular elastances are denoted $E_\text{max}$  and $E_\text{min}$. $E_\text{max}$ is the end-systolic elastance and $E_\text{min}$ is the end-diastolic elastance. The functional form of the elastance $E(t)$ is given during the time interval $[0, T]$ as follows:
\begin{equation}
    E(t) =  k \left(\frac{g_1(t)}{1+g_1(t)}\right)\left(\frac{1}{1+g_2(t)}-\frac{1}{1+g_2(T)}\right) + E_\text{min},
    \label{eq:elastance}
\end{equation}
where
\begin{equation}
    g_1(t) = \left(\frac{t}{\tau_1}\right)^{m_1}, \quad g_2(t) = \left(\frac{t}{\tau_2}\right)^{m_2}.
\end{equation}
Here $T$ is the period of the heartbeat. Note that equation \eqref{eq:elastance} makes $E(0)$ = $E(T)$. (We have made a slight modification of the formula used in \cite{Mynard12} to ensure this.) Outside of the interval $[0, T]$, we define $E(t)$ as a periodic function with period $T$, so that $E(t)$ = $E(t + T)$ for all $t$. The parameter $k$ is chosen so that the maximum value of $E(t)$ is $E_\text{max}$. The formula for $k$ to achieve this is 
\begin{equation}
    k = \frac{E_\text{max}-E_\text{min}}{\max_{t\in [0,T]}[(\frac{g_1(t)}{1+g_1(t)})(\frac{1}{1+g_2(t)}-\frac{1}{1+g_2(T)})]}
\end{equation}
The maximum in the denominator of the formula for $k$ is computed by evaluating the expression that needs to be maximized at a collection of equally spaced points within the interval $[0, T]$, and then choosing the largest of the values of that expression that are found. Although this procedure does not yield the exact maximum value, it comes close enough for practical purposes, especially since the goal is to find the maximum value, rather than the time at which it occurs. 
Parameter values used for the heart chambers are provided in Table \ref{table:elastance}. When two values are shown, the first corresponds to the normal model and the second corresponds to the pre-intervention RPH model. The constant $\tau_{1}$ controls the timescale of contraction, $\tau_{2}$ controls the duration of systole, and $m_1$ and $m_2$ govern the speed of contraction and relaxation respectively. Note that $\tau$ and $m$ are estimated from previously employed values \cite{Stergiopulos96}, and the values for $E_\text{min}$ and $E_\text{max}$ are similar to those used by \cite{liang2009multi}. 

\begin{table}[!b]
\begin{center}
\begin{tabular}{l*{6}{c}r}
\hline
Parameters              & Symbol & Units & Left Ventricle & Right Ventricle \\
\hline
Minimum elastance & $E_\text{min}$ & mmHg/L & $0.08\times 10^3$ & $0.04\times 10^3$,\, $0.08\times 10^3$  \\
Maximum elastance            & $E_\text{max}$ & mmHg/L &  $3.00\times 10^3$ & $0.60 \times 10^3$,\, $3.00\times10^{3}$  \\
Contraction exponent           & $m_1$ & - & 1.32 & 1.32  \\
Relaxation exponent     & $m_2$ & - & 27.4 & 27.4  \\
Systolic time constant     & $\tau_1$ & min & 0.269$\times T$ & 0.269$\times T$  \\
Diastolic time constant     & $\tau_2$ & min & 0.452$\times T$ & 0.452$\times T$ \\
Dead Volume     & $V_d$ & L & 0.010 & 0.010 \\
Period of heartbeat     & $T$ & min & 0.0125 & 0.0125 \\
\end{tabular}
\end{center}
\caption{Parameters for the time varying compliances in the heart model. When two numbers are given, the first one is used in the normal model and the second one is used in the RPH model. The parameters shown in the table are resting values. The period of a cardiac cycle, $T$, will be modified in the exercise model as described in section \ref{sec:exercise_models}.}
\label{table:elastance}
\end{table}

\subsection{Shunt model}
\label{subsec:shunt}
The Gorlin equation is used to calculate flows through surgically created shunts (ASD, VSD, or Potts shunt) in our model \cite{reynolds1990determination}. This allows us to specify the cross-sectional area of the connection that the surgeon creates, and to study how the shunt size affects hemodynamic variables and the transport of oxygen.

To derive the shunt model, consider two chambers, denoted by the indices 1 and 2, separated by a wall with a hole in it that corresponds to the shunt.  Let $A_0$ be the cross-sectional area of the hole. We assume that the velocity of the blood as it goes through the hole is much larger than the velocity in the two chambers, so that we may consider the fluid in each of the two chambers as if it were at rest. Let $Q$ denote the volume of blood flow per unit time through the hole, with the direction from chamber 1 to chamber 2 considered positive. Then, the spatially averaged velocity of blood flow in the hole itself is given by
\begin{equation}
    v = Q / A_0.
\end{equation}
Let $P_1$ and $P_2$ be the pressures in the two chambers, and let $P_0$ be the pressure within the hole.
Suppose, for example, that $Q$ $>$ 0. By Bernoulli's equation in the upstream chamber up to the hole itself, one has
\begin{align}
    P_0 & = P_1 - \frac{1}{2}\rho v^2 = P_1 - \frac{\rho}{2A_0^2}Q^2.
\end{align}
In the region downstream of the hole, Bernoulli's equation does not apply because the flow there is dominated by turbulent eddies that dissipate energy. The result is that the pressure is relatively constant in the downstream region, in particular that $P_2 = P_0$.
It follows that
\begin{equation}
    P_1 - P_2 = \frac{\rho}{2A_0^2}Q^2, \quad  Q > 0.
\label{eq:gorlin1}
\end{equation}
By the same reasoning, for flow in the other direction
\begin{equation}
    P_2 - P_1 = \frac{\rho}{2A_0^2}Q^2, \quad  Q < 0.
\label{eq:gorlin2}
\end{equation}
Equations \eqref{eq:gorlin1} and \eqref{eq:gorlin2} can be combined as follows:
\begin{equation}
    P_1 - P_2 = \frac{\rho}{2A_0^2}|Q|Q,
\end{equation}
This shows that the hydraulic resistance of the hole is given by
\begin{equation}
    R_\text{shunt} = \frac{\rho}{2A_0^2}|Q|.
\end{equation}
Note that the above formula for $R_\text{shunt}$ is independent of viscosity. In reality, there is a very small viscous resistance as well, so we modify the above formula to read
\begin{equation}
    R_\text{shunt} = R_\text{visc} + \frac{\rho}{2A_0^2}|Q|.
\label{eq:gorlin3}
\end{equation}
In most situations $R_\text{visc}$ is negligible, but we should include it to prevent $R_\text{shunt}$ from being zero, which would otherwise happen in principle every time that $Q$ changes sign. Since $R_\text{visc}$ is included only for this reason, we choose the very small value $R_\text{visc}$ = 0.1 mmHg/(L/min). Evaluating the conductance of the hole from equation \eqref{eq:gorlin3}, one obtains
\begin{equation}
    G_\text{shunt} = \frac{1}{ R_\text{visc} + \frac{\rho}{2A_0^2}|Q| }
\label{eq:gorlin4}
\end{equation}
In making use of equation \eqref{eq:gorlin4}, one must be careful about units. In this paper, we use what may be called physiological units, in which volume is measured in liters, pressure in mmHg, and time in minutes. The constants $A_0$ and $\rho$ need to be expressed in these units. The use of liters for volume implies that our unit of length is the decimeter (dm), which is equal to 10 cm. Thus, $A_0$ needs to be expressed internally in terms of dm$^2$, although in the presentation of our results, we use cm$^2$ since these units have more meaning to the reader. To express density in physiological units, one needs the units of mass. The units of force are mmHg $\cdot$ dm$^2$, and the units of acceleration are dm/min$^2$, so the units of mass are mmHg $\cdot$ dm $\cdot$ min$^2$. Dividing this by dm$^3$, one obtains the units of density as mmHg $\cdot$ (min/dm)$^2$. After taking units carefully into account in this way, the density in physiological units is
\begin{align}
    \rho = 0.00002084167\cdot\frac{\text{mmHg} \cdot {\text{min}^2}}{\text{dm}^2}.
\label{eq:gorlin5}
\end{align}



\begin{figure}[h!]
    \centering
    \includegraphics[width=0.48\textwidth]{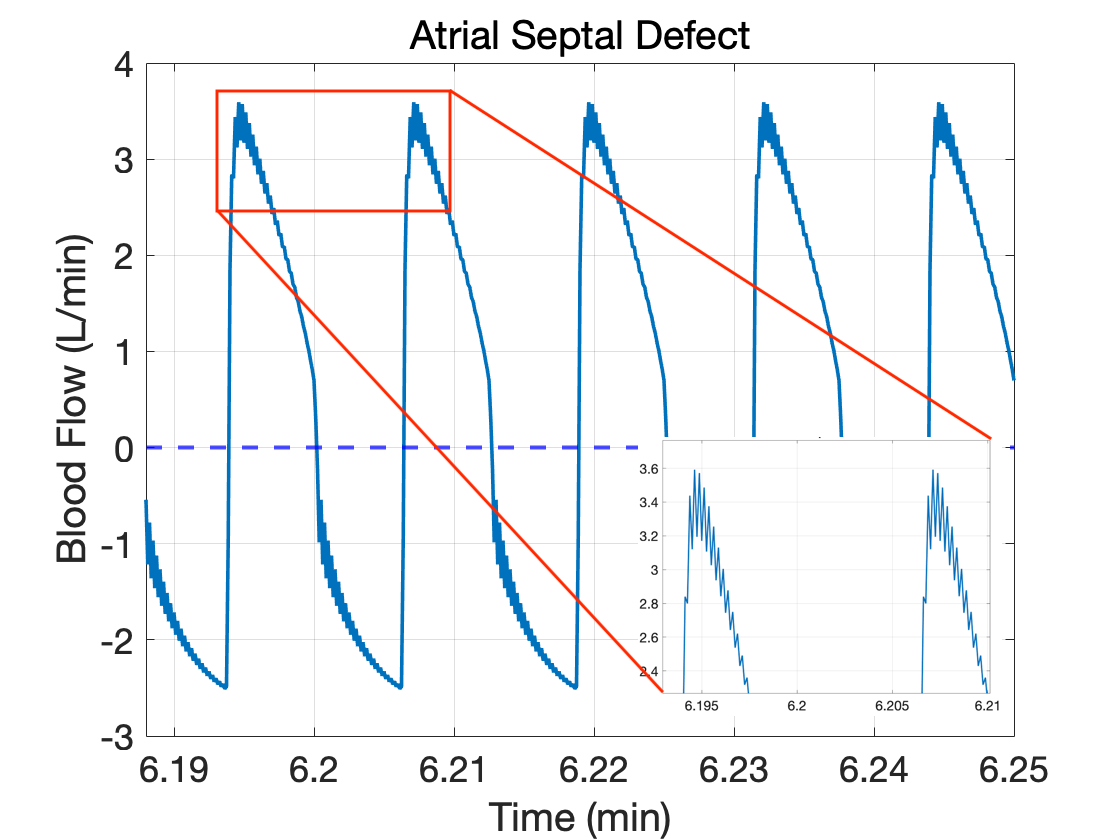}
    \includegraphics[width=0.48\textwidth]{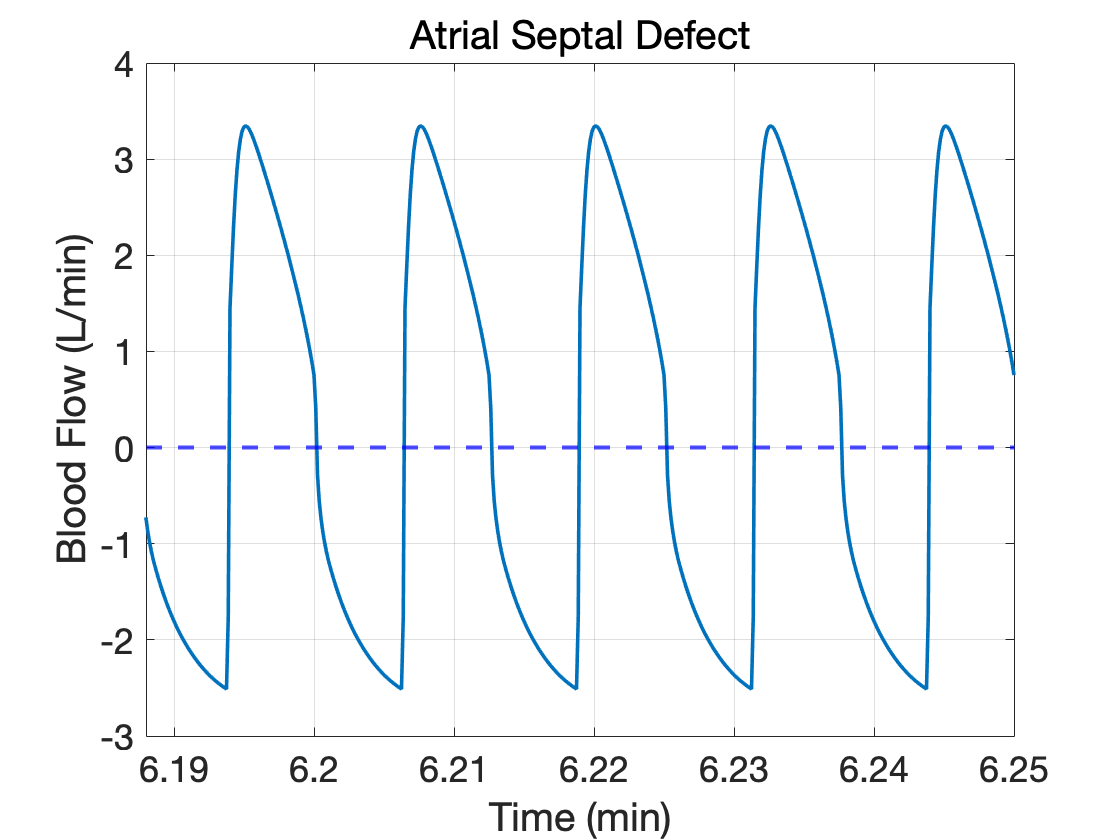}
     \caption{A comparison of shunt flow waveforms from the ASD model, obtained with and without fixed-point iterations (due to the nonlinearity in the shunt conductance). The shunt area is 1 $\text{cm}^2$. The left panel shows flow computed without the fixed-point iteration. The right panel shows flow computed with the fixed-point iteration. Notice that the fixed-point iteration removes the high-frequency oscillations.}
    \label{fig:fixedpoint}
\end{figure}

Another complication in the use of equation \eqref{eq:gorlin4} is that the shunt conductance $G_\text{shunt}$ is flow-dependent. A simple idea here would be to use the shunt flow on the previous time step to set the shunt conductance for the present time step, but instead of this, we use a fixed-point iteration, with the shunt flow on the previous time step as the initial guess. At each step of the fixed-point iteration, equation \eqref{eq:gorlin4} is used to set the shunt conductance based on the latest guess for the shunt flow. Then, the shunt conductance is inserted into the appropriate two places in the conductance matrix $G$ (one entry for each flow direction, since there is no valve involved in the shunt). Finally, all of the pressures and flows for the circulation are computed, including the shunt flow. The benefit of doing the fixed-point iteration can be seen in Figure \ref{fig:fixedpoint} since it removes the numerical oscillations seen in the blood flow waveform. In practice, 10 fixed-point iterations are used for each time step, and this achieves good enough agreement between the flow that is used to set the shunt conductance and the flow that is calculated on the basis of that shunt conductance. 

\subsection{Oxygen transport model}
\label{subsec:oxygenmodel}

An important consequence of the surgical interventions considered in this paper is the mixing  of oxygenated and doxygenated blood. Our approach to the modeling of oxygen transport follows Tu and Peskin \cite{Tu89}. Time-varying oxygen concentrations for each compliance chamber are described by the following system of differential equations: 
\begin{equation}
\label{eq:o2one}
\frac{d}{dt}([O_{2}]_{i}V_{i}) = \displaystyle\sum_{j = 1\atop j\ne i}^{N}([O_{2}]_{j}Q_{ji} - [O_{2}]_{i}Q_{ij} + M_{ji}).
\end{equation}
The variable $[O_{2}]_{i}$ is the oxygen concentration in compliance chamber $i$, the variable $Q_{ji} $ is the blood flow from $j$ to $i$, and the parameter $M_{ji}$ is the rate at which oxygen is added to the stream of blood that is flowing from chamber $j$ to chamber $i$. Note that $M_{ji}$ is positive if oxygen is being added to the blood stream, and negative if oxygen is being removed. The correctness of equation \eqref{eq:o2one} relies on the fact that all of the flows that appear in it are positive or zero. This is a benefit of our formulation in which every connection is equipped with a valve, as described above in subsection \ref{subsec:hemo}. These equations describe conservation of oxygen during transport between chambers, metabolic consumption of oxygen within systemic organs, and replenishment of oxygen within the lungs. After computing the flows at time step $n$,  those flow values are used to update the oxygen concentrations from time step $n-1$ to $n$ as follows:
\begin{equation} 
\label{eq:o2two} 
\frac{[O_{2}]_{i}^nV_{i}^n - [O_{2}]_{i}^{n-1}V_{i}^{n-1}}{\Delta t} = \displaystyle\sum_{j = 1 \atop j\ne i}^N ([O_{2}]_{j}^{n-1} Q_{ji}^n -[O_{2}]_{i}^{n-1}Q_{ij}^n + M_{ji}).
\end{equation}

Note that this is the forward Euler method insofar as the oxygen concentrations are concerned, although it differs from the forward Euler method by using the flows at time step $n$. The manner in which $M_{ji}$ is determined for use in these equations is described below.

We use the millimole (mmol) as the unit for the amount of oxygen. It then follows from our other choices of units that the units of oxygen concentration are mmol/L and the units of the rate of oxygen consumption by the body are mmol/min. A standard concentration of hemoglobin in blood is 2.5 mmol/L, and since each hemoglobin molecule can carry four oxygen molecules, the oxygen concentration when hemoglobin is fully saturated is 10 mmol/L.

There are only two places in our model where the parameter $M_{ji}$ that appears in equation \eqref{eq:o2one} is nonzero. One of these is in the connection from the pulmonary arteries (pa) to the pulmonary veins (pv). We assume that $M_\text{pa,pv}$ is such that the stream of blood flowing from the pulmonary arteries to the pulmonary veins becomes fully saturated with oxygen during its passage through the pulmonary capillaries. This gives the equation
\begin{equation}
\label{eq:o2three} 
M_\text{pa,pv} = (10 \text{ mmol/L} - [O_2]_\text{pa})\, Q_\text{pa,pv}.
\end{equation}
Equation \eqref{eq:o2three} is used to set $M_\text{pa,pv}$ at every time step. Note that this is \textit{not} the same as setting $[O_2]_\text{pv}$ = 10 mmol/L. The reason for this is that there may be other streams of blood entering the pulmonary venous compartment besides the one coming from the pulmonary arteries. In particular, since we regard the left atrium as being part of the pulmonary venous compartment, this will be the case when we are simulating a surgically created atrial septal defect.

The other nonzero value of $M_{ji}$ in our model is $M_\text{sa,sv}$, which is negative, since it represents oxygen consumption by the tissues. This oxygen is extracted from the stream of blood that flows from the systemic arteries (sa) to the systemic veins (sv). In the simulations reported here, we keep $M_\text{sa,sv}$ constant, and we calculate its value by using a normal cardiac output of 5.6 L/min and a normal amount of oxygen extraction by the systemic tissue of 30\%. These values result in the following:
\begin{align}
\label{eq:o2four} 
- M_\text{sa,sv} = 0.3 \cdot (10 \text{ mmol/L}) \cdot (5.6 \text{ L/min}) & = 16.8 \text{ mmol/min}
\end{align}
The initial value for the oxygen concentration is set to 10 mmol/L in all compartments, but, this has no effect on our results because we run the simulations to a periodic steady state.


\subsection{Exercise model}
\label{sec:exercise_models}

The surgical interventions studied in this paper have an impact on exercise tolerance. We develop a simple exercise model to use within our hemodynamic and oxygen transport models in order to compare exercise tolerance pre and post intervention \cite{inbook, article1}. The independent variable in our exercise model is oxygen consumption, denoted $M$, in the systemic tissues. This independent variable is used to set the parameter $M_\text{sa,sv} = -M$ in equations \eqref{eq:o2two} and determine several other parameters as described below. The {\em resting} oxygen consumption in the systemic tissues is denoted $M_\text{rest}$ which we set equal to 16.8 mmol/min. The heart rate is taken to be a function of the oxygen consumption using the following equation, adapted from \cite{article}: 
\begin{align}
\label{eq:exone} 
\emph{\text{HR}} = (0.94 \text{ beats/mmol}) \times (M - M_\text{rest}) + 80 \text{ beats/min}.
\end{align}
When changing the heart rate, the period $T = 1/\text{\em HR}$ is modified accordingly. Note that to create the resting condition, we set $M = M_\text{rest}$, in which case we recover the resting heart rate of 80 beats/min. The systemic resistance $R_S$ decreases during exercise and is taken to be a function of the heart rate as follows:
\begin{align}
\label{eq:extwo} 
R_{S} (\text{\em HR}) = \frac{(17.5 \text{ mmHg/(L/min)}) \times (80 \text{ beats/min})}{\emph{\text{HR}}}. 
\end{align}
Note that in the resting case, {\em HR} = 80 beats/min, and we recover the resting systemic resistance of 17.5 mmHg/(L/min). The third component of our exercise model is a decrease in the systemic venous dead volume. This modification is described by the following formula:
\begin{align}
\label{eq:exthree} 
V_{d,SV}(\text{\em HR}) = (3.1 \text{ L}) \times \left( \frac{80 \text{ beats/min}}{\emph{\text{HR}}}\right)^{0.1}, 
\end{align}
where we recover the resting dead volume shown in Table \ref{table:parms1} when $\text{\em HR} = 80$ beats/min. Here, it is better to think of the dead volume as a \textit{reserve volume} that can be mobilized by the sympathetic nervous system by constricting the systemic veins.  This has the effect of supporting and even moderately increasing the stroke volume of the heart, despite the increasing heart rate associated with exercise.

\section{Results and discussion}
\label{sec:results}

In this section, we examine the simulated effects for each of the three interventions. First, we compare results from the normal and pre-intervention RPH models. Second, we simulate each surgical intervention within our models in order to investigate changes in pressure and oxygen saturation in the systemic and pulmonary arteries. Third, we consider various levels of exercise in a normal model, a pre-intervention RPH model, and in two post-interventions RPH models, the VSD and Potts shunt.  Oxygen saturation, systemic flow, and oxygen delivery are studied as the oxygen consumption in the systemic tissues is varied. For the post-intervention cases, we focus on a shunt size that most significantly lowers the pulmonary artery pressure in both the VSD and Potts shunt. Fourth, we compare the resting model with the exercise model. In these cases, we study systemic flow, oxygen saturation, and oxygen delivery as the pulmonary resistance is varied because of favorable post-intervention remodeling of the pulmonary vasculature. Finally, we examine mean shunt flow and shunt flow waveforms to determine whether the shunt is indeed right-to-left, as anticipated, or is perhaps bidirectional. 

In all simulations, 100 time steps are used for each cardiac cycle. The heart rate at rest is 80 beats/min. Each computer experiment for both rest and exercise is run for 500 cardiac cycles. This duration is sufficient for each simulation to achieve a periodic steady state for all variables in all cases. We remark that a periodic steady state for the hemodynamic variables is achieved very quickly, within 10-20 cardiac cycles. In contrast, it takes many more cardiac cycles for the oxygen concentrations to reach a periodic steady state. When we report a single value for any quantity as the result of a simulation, it is the average of that quantity over the last five cardiac cycles.

\begin{table}
\begin{center}
\begin{tabular}{l*{6}{c}r}
\hline
Model Output  & Units & Normal & RPH Pre-Intervention \\
\hline
Total blood volume & L & 5.098 & 5.248 \\
Cardiac output            & L/min & 5.589 & 3.661\\
Left ventricle stroke volume     & mL & 69.871 & 45.762 \\
Right ventricle stroke volume     & mL & 69.871 &45.762 \\
Diastolic systemic arterial pressure     & mmHg & 80.844 & 56.109  \\
Systolic systemic arterial pressure     & mmHg & 118.561 & 81.077  \\
Mean systemic arterial pressure     & mmHg & 102.249 & 70.327 \\
Diastolic pulmonary arterial pressure      & mmHg & 13.834 & 80.635 \\
Systolic pulmonary arterial pressure      & mmHg & 24.143 & 96.224  \\
Mean pulmonary arterial pressure      & mmHg & 19.505 & 89.526\\
\end{tabular}
\end{center}
\caption{Comparison of model outputs in normal and pre-intervention RPH circulations. The parameters that are changed to convert the normal circulation to the RPH circulation are described in Subsection \ref{subsec:hemo}.}
\label{table:clinical}
\end{table}

\subsection{Pressures and oxygen saturations }
\label{sec:pressure_oxygen}

First, we compare the pre-intervention RPH model to the normal model. Values for different hemodynamic variables are shown in Table \ref{table:clinical}. As discussed above, parameters for the pre-intervention RPH model are derived from the normal model as follows. The pulmonary resistance is taken to be 1.3 times the systemic resistance (to describe the onset of pulmonary vascular disease), the right ventricular elastances are taken to be the same as the left ventricular elastances (to describe right-heart remodeling in the setting of increased afterload), and the blood volume is increased from 5.098 L to 5.248 L (to describe compensation by the body to increase cardiac output in the setting of heart failure). These changes result in a pre-intervention RPH model with pulmonary pressures that exceed systemic pressures. This physiologic feature is seen clinically and motivates the need for the types of surgical interventions explored in this paper. As expected, cardiac output and stroke volume are substantially lower in the RPH model as compared to the normal model.

\begin{figure}[hbt!]
\minipage{0.48\textwidth}
  \includegraphics[width=\linewidth]{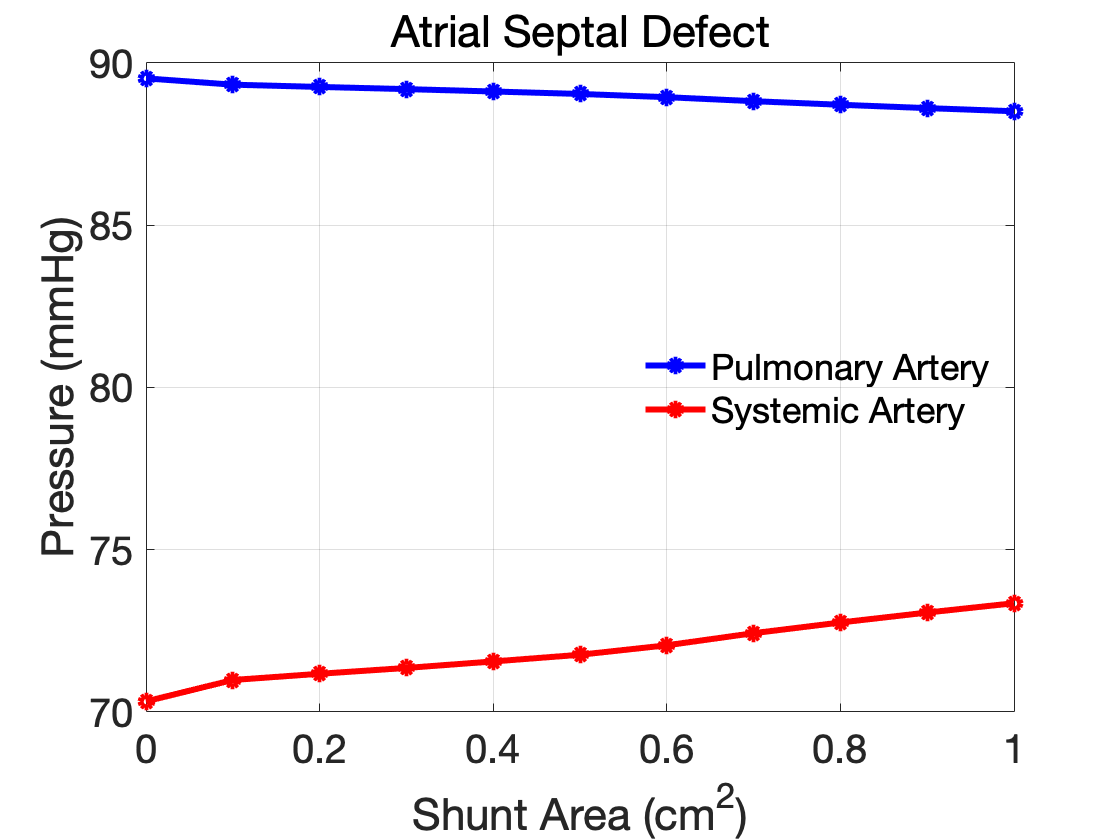}
\endminipage\hfill
\minipage{0.48\textwidth}
  \includegraphics[width=\linewidth]{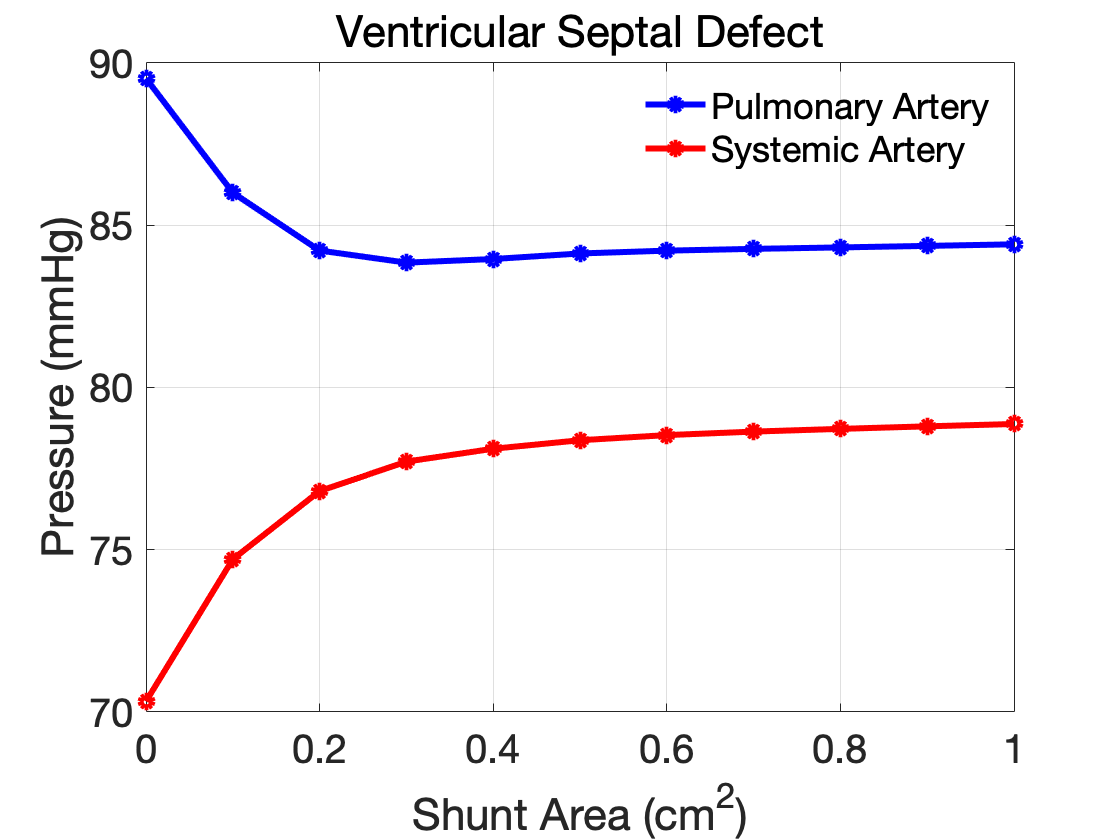}
\endminipage\hfill
\minipage{0.48\textwidth}%
  \includegraphics[width=\linewidth]{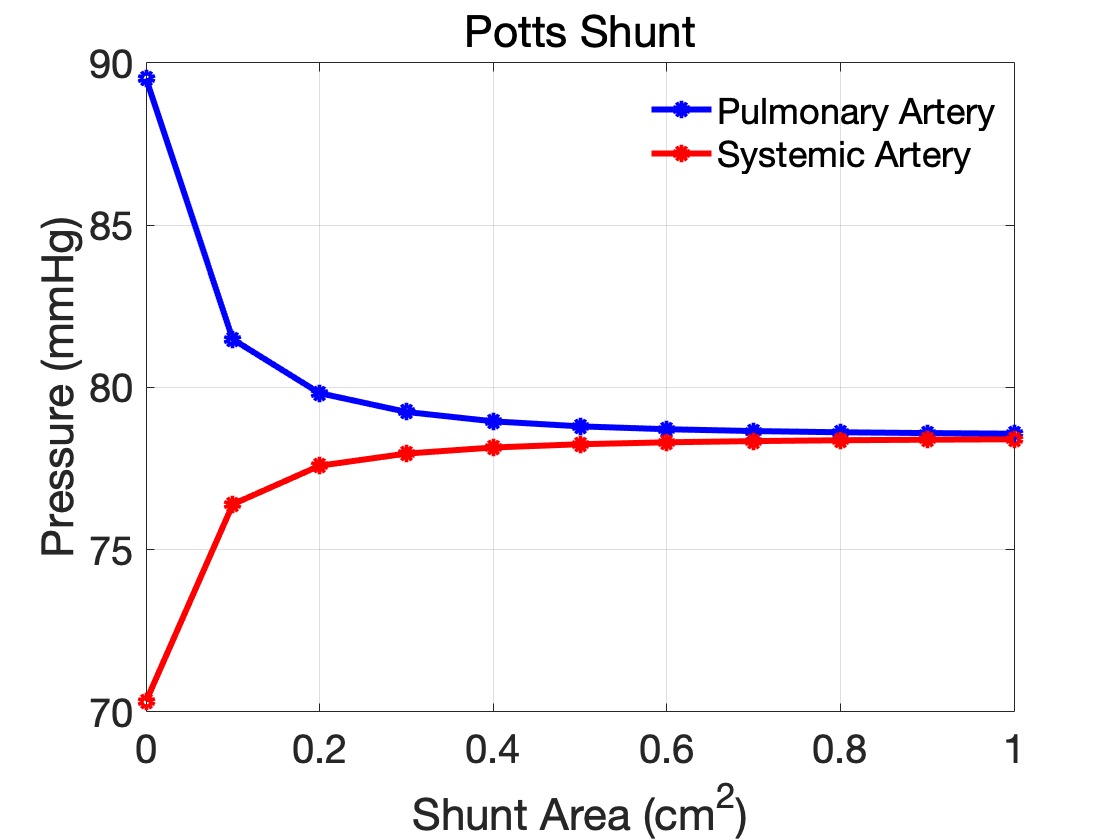}
\endminipage
     \caption{A comparison of mean pressures in the pulmonary artery (blue) and systemic artery (red) as the shunt area is varied in the ASD, VSD, and Potts shunt.}
    \label{fig:pressures}
\end{figure}

\begin{figure}[hbt!]
    \centering
    \includegraphics[width=0.48\textwidth]{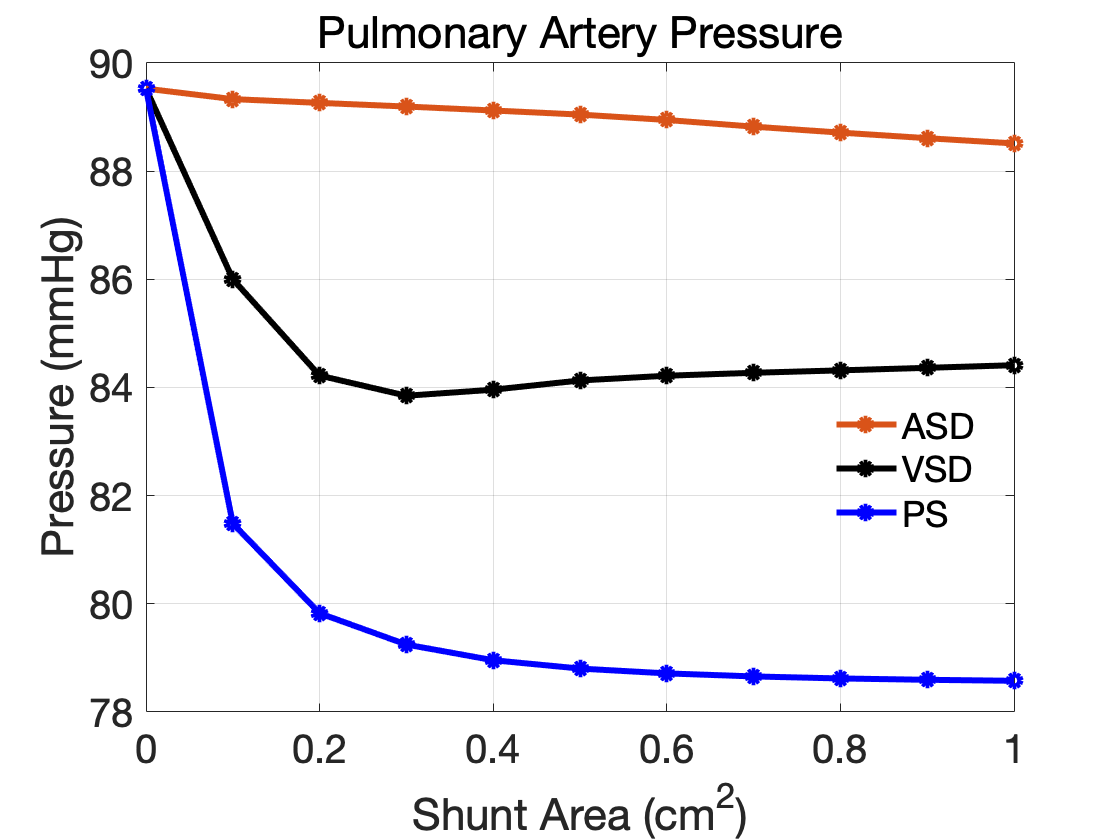}
    \includegraphics[width=0.48\textwidth]{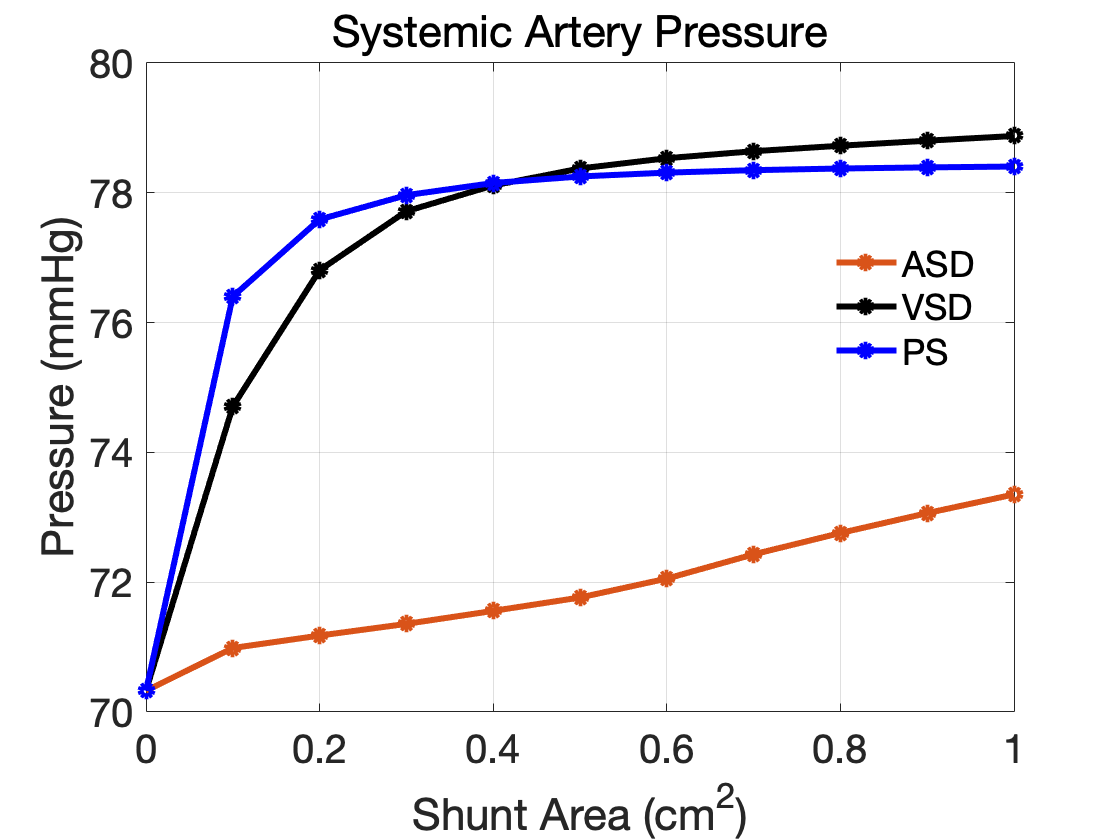}
     \caption{A comparison of mean pressures in the pulmonary artery and the systemic artery for the three interventions: the left panel shows the pulmonary artery pressures for the ASD, VSD, and Potts shunt as the shunt area is varied. The right panel shows the systemic artery pressures for the three interventions as the shunt area is varied. Note different pressure scales in the two panels.}
    \label{fig:pasa}
\end{figure}

Next, we consider the impact of each intervention on pressures and oxygen saturations. Figure \ref{fig:pressures} shows the systemic and pulmonary arterial blood pressures (mean values) as functions of the shunt area. Results for the atrial septal defect (ASD) are in the left panel, results for the ventricular septal defect (VSD) are in the right panel, and results for the Potts shunt are in the bottom panel. The ASD results show that this intervention is not successful in lowering the pulmonary arterial pressure, which is perhaps consistent with the fact that an ASD is a volume-unloading shunt. There appears to be a very small effect from the ASD, but it is certainly not one that would be therapeutic. The VSD intervention lowers the mean pulmonary artery pressure from about 90 mmHg to about 84 mmHg, which could be beneficial. This smallest shunt for which this result is achieved has cross-sectional area equal to 0.3 cm$^2$. It is interesting to note that beyond this value for the shunt area, the pulmonary arterial pressure slightly increases as the shunt size increases. The Potts shunt most substantially lowers the mean pulmonary arterial pressure, from about 90 mmHg to about 78 mmHg. Unlike in the VSD case, the pulmonary arterial pressure with the Potts shunt decreases monotonically with increasing shunt size, but most of the benefit has already occurred with a shunt size of 0.3 cm$^2$. Our model suggests that there is little benefit in using a larger Potts shunt size than this value. Figure \ref{fig:pasa} shows the pressures in the pulmonary artery and in the systemic artery for the three interventions, all on the same plot, as functions of the shunt area.

\begin{figure}[h!]
\minipage{0.48\textwidth}
  \includegraphics[width=\linewidth]{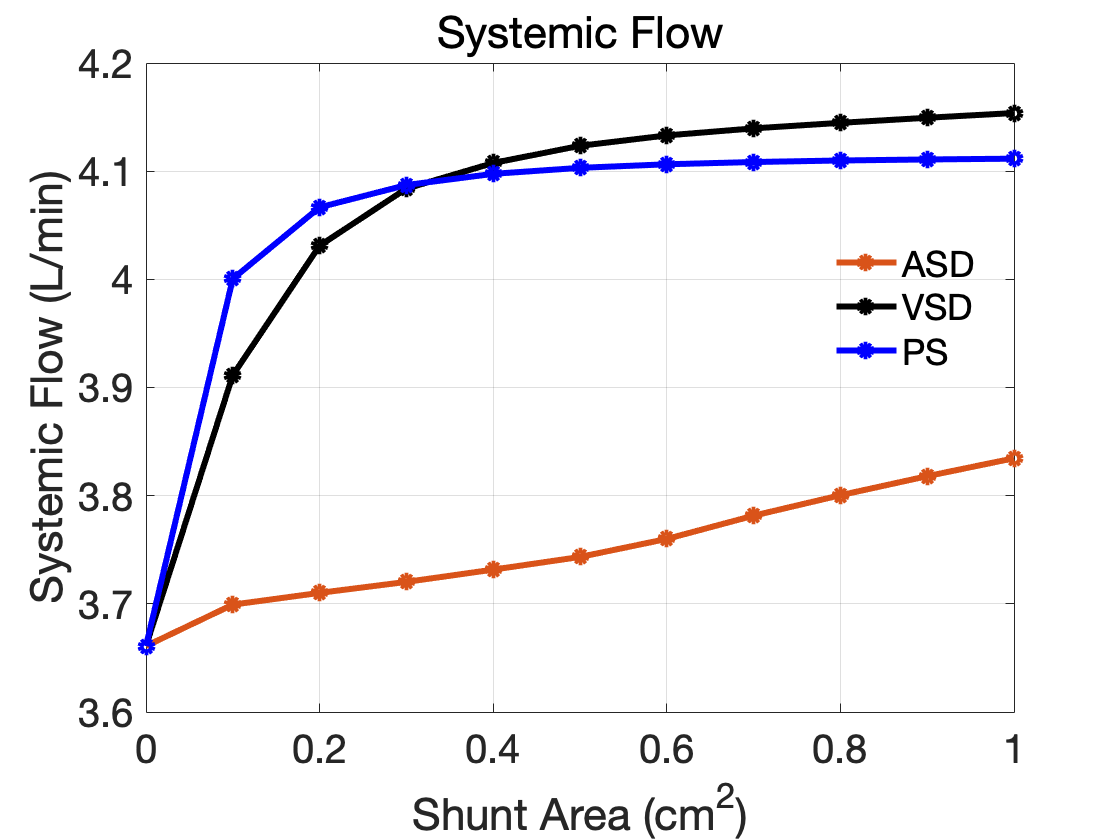}
\endminipage\hfill
\minipage{0.48\textwidth}
  \includegraphics[width=\linewidth]{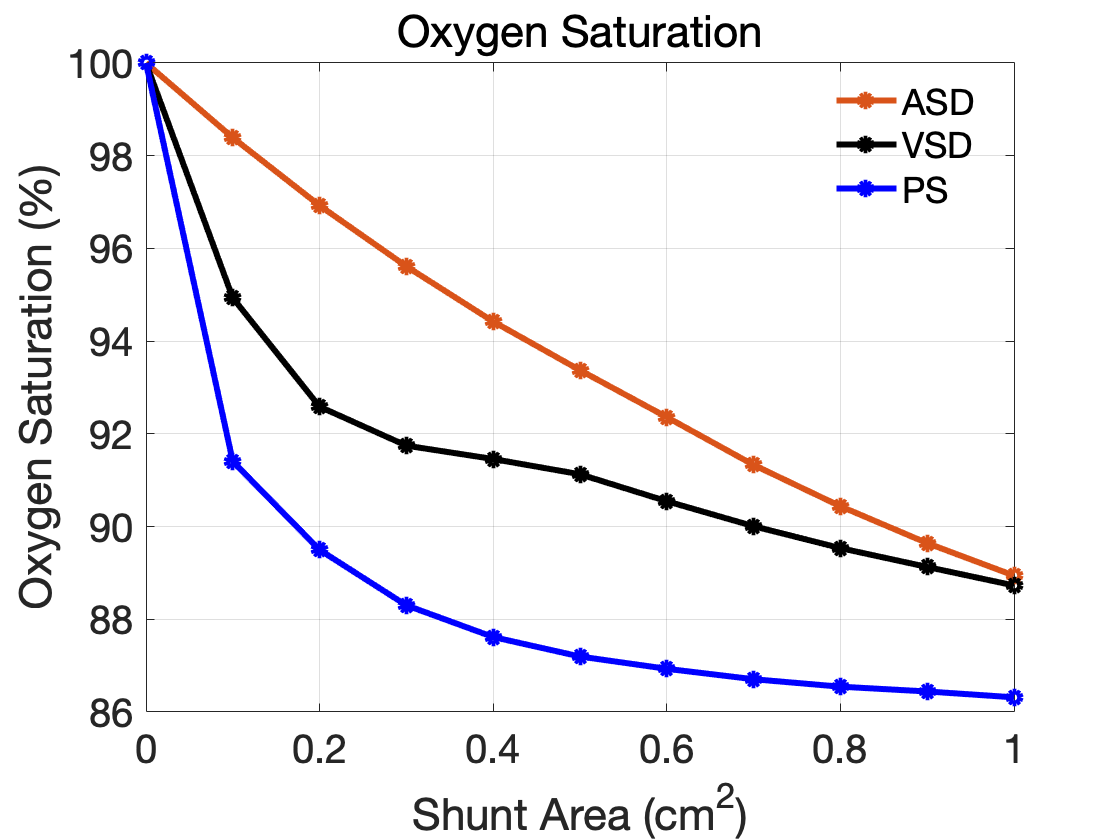}
\endminipage\hfill
\minipage{0.48\textwidth}%
  \includegraphics[width=\linewidth]{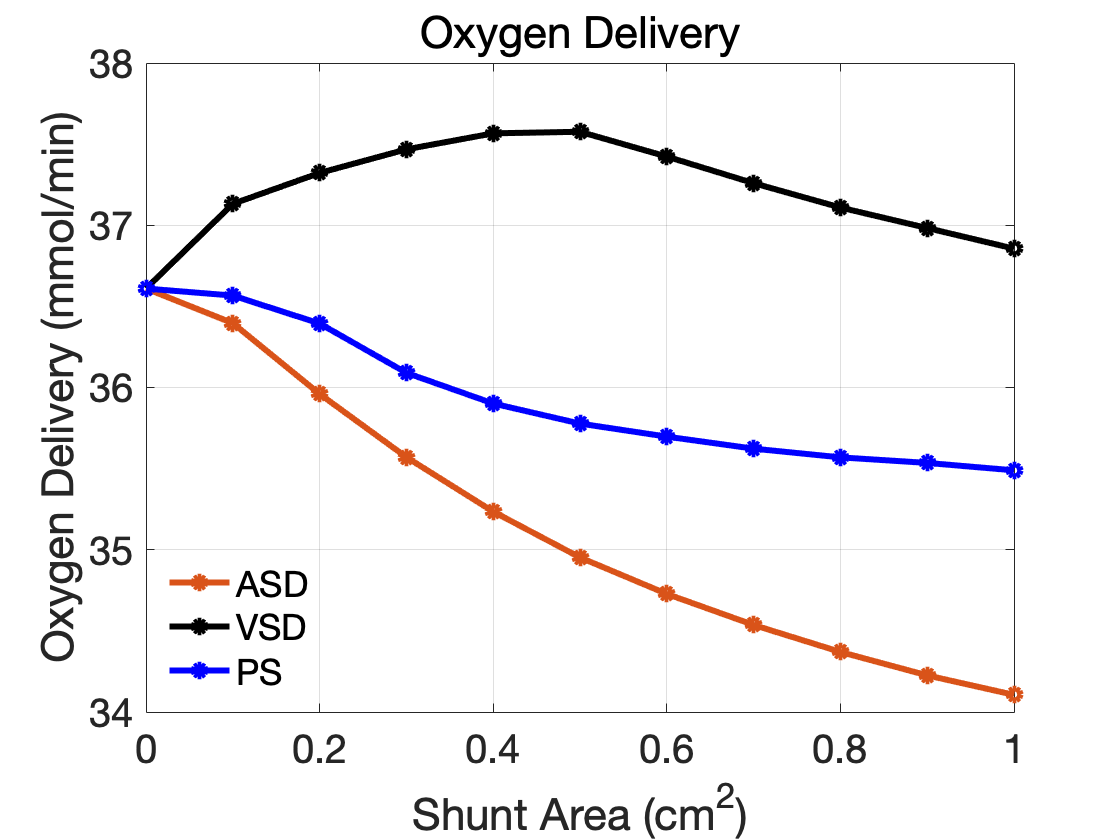}
\endminipage
   \caption{A comparison of systemic flow, oxygen saturation, and oxygen delivery rate for the ASD, VSD, and Potts shunt.}
\label{fig:o2one}
\end{figure}

We further investigate oxygen transport for each intervention in Figure \ref{fig:o2one}. This figure depicts systemic flow, oxygen saturation, and oxygen delivery for the three interventions. Systemic flow is relevant here because it is used in the computation of oxygen delivery to the systemic tissues. Oxygen saturation is the oxygen concentration (in mmol/L) expressed as a percentage of 10 mmol/L, which is the maximum possible oxygen concentration in our model. The rate at which oxygen is delivered to the systemic tissues is calculated by multiplying the systemic flow by the systemic arterial oxygen concentration. 

All three interventions increase systemic flow. The increase is substantial in the case of VSD and Potts shunt. This increase in systemic flow has an important effect on oxygen delivery to the systemic tissues. All three interventions decrease systemic arterial oxygen saturation. This is inevitable, since the interventions by design are allowing deoxygenated blood to bypass the lungs. The effect is smallest in the case of the ASD, but since this intervention has minimal benefit in terms of decreasing pulmonary artery pressure, the fact that it also does the least harm is not really of interest. The VSD and Potts shunt produce similar decreases in systemic arterial oxygen saturation, but these two interventions look quite different from the point of view of oxygen delivery to the systemic tissues. The increase in systemic flow in the VSD case seems to compensate nicely for the drop in systemic arterial oxygen saturation. At the shunt size of about 0.5 cm$^2$, the oxygen delivery increases by about 3\% compared to the delivery in the pre-intervention state (corresponding to a shunt size of zero). Recall, however, that the optimal reduction in pulmonary artery pressure occurs in the VSD at a shunt size of about 0.3 cm$^2$, and at this size, the increase in delivery is smaller.

\subsection{Exercise tolerance}

\begin{figure}[h!]
\minipage{0.48\textwidth}
  \includegraphics[width=\linewidth]{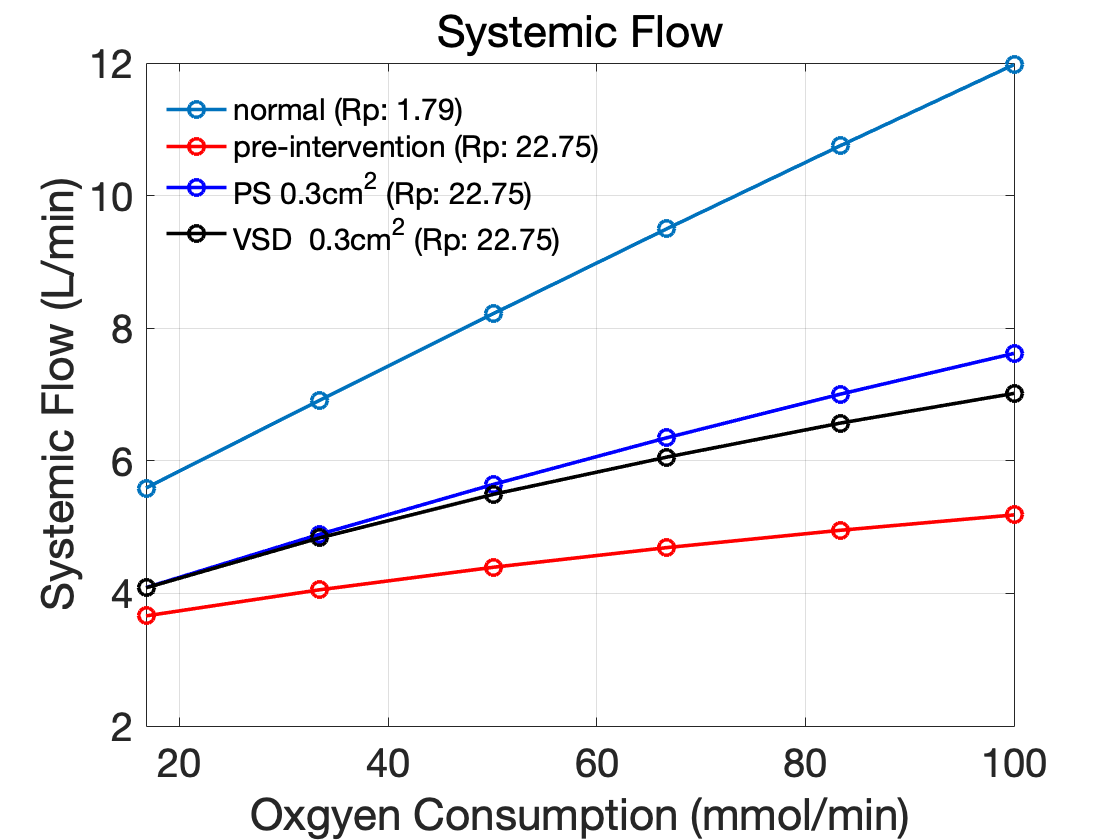}
\endminipage\hfill
\minipage{0.48\textwidth}
  \includegraphics[width=\linewidth]{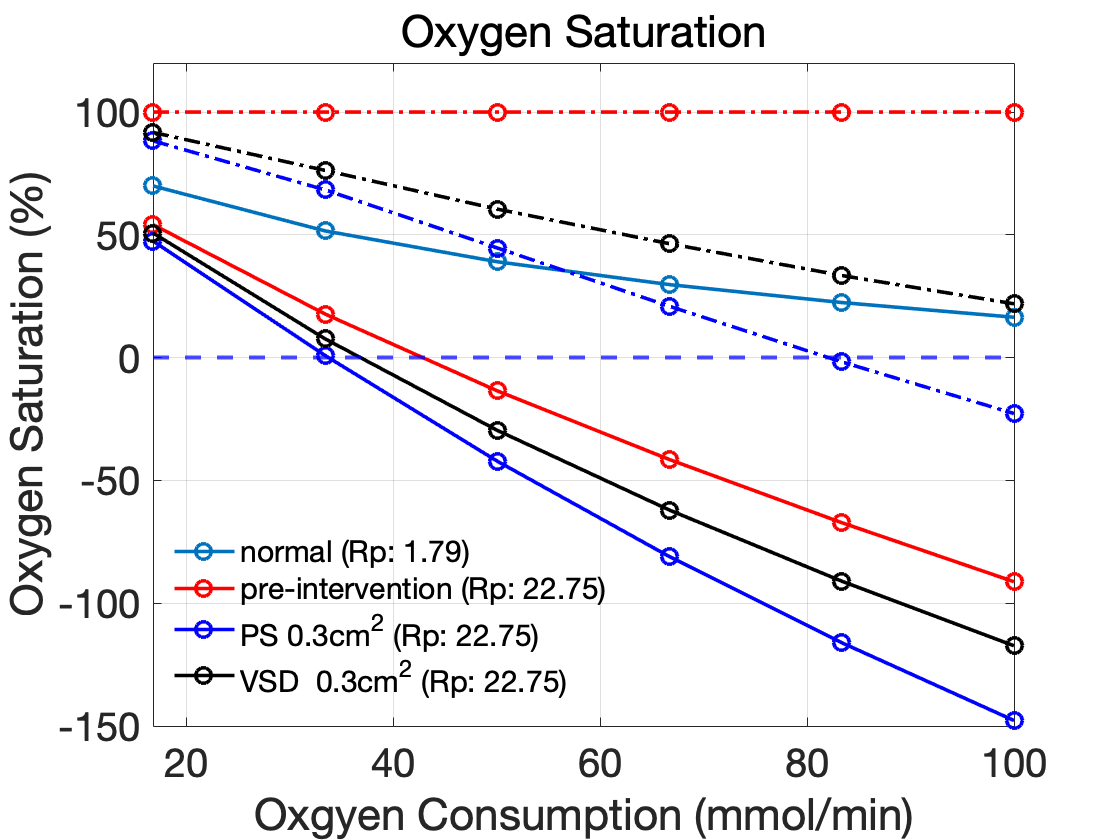}
\endminipage\hfill
\minipage{0.48\textwidth}%
  \includegraphics[width=\linewidth]{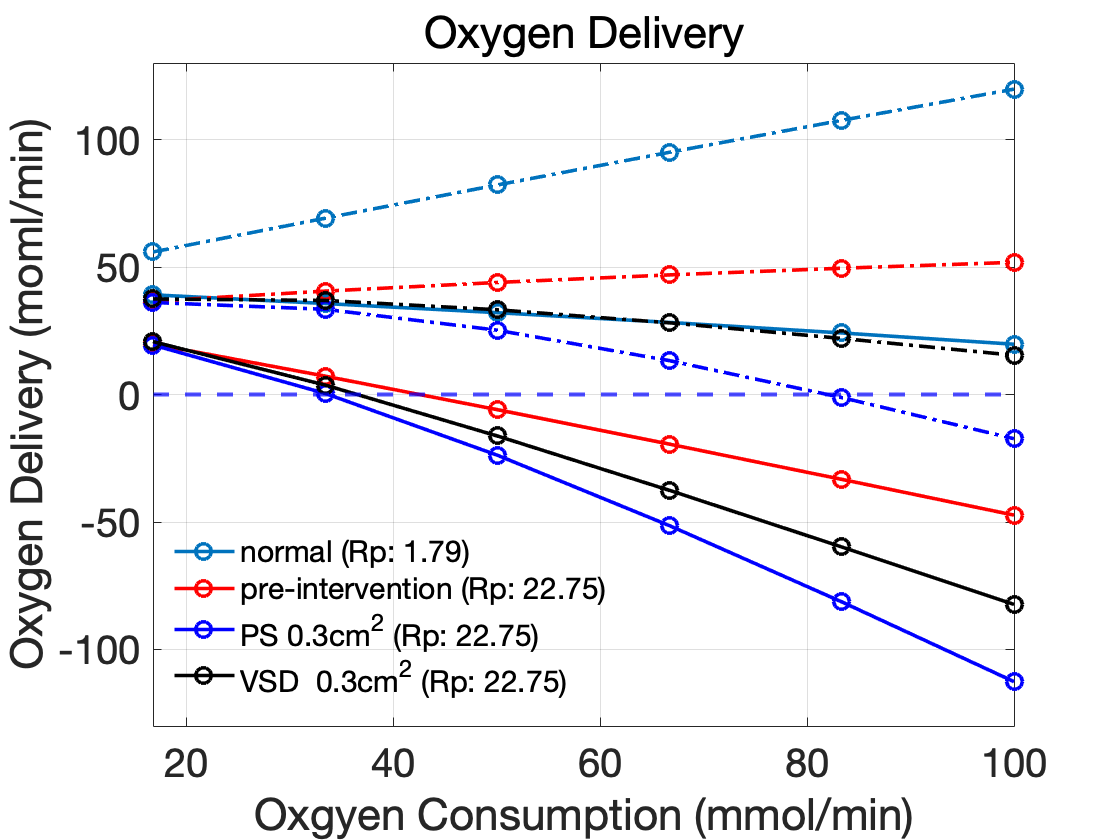}
\endminipage
   \caption{A comparison of systemic flow, oxygen saturation, and oxygen delivery as functions of oxygen consumption, which corresponds to the level of exercise and is the independent variable in our exercise model. Only the VSD and Potts shunt are considered. If the computed value of the systemic venous saturation is negative, this implies the assumed level of oxygen consumption and the corresponding exercise level is not possible. In the right and bottom panels, the dashed-dotted lines correspond to the systemic arteries, and the solid lines correspond to the systemic veins. $R_P$ is the pulmonary resistance in units of mmHg/(L/min).}
\label{fig:exercise1}
\end{figure}

In this section, we consider the effect of exercise in four cases: (1) the normal circulation, (2) the pre-intervention circulation with refractory pulmonary hypertension, and a post-intervention circulation with either a (3) VSD or (4) Potts shunt. The ASD intervention is not considered here since it does not appear to be useful in lowering pulmonary artery pressure. The normal circulation is considered as a point of reference. We consider a shunt area of 0.3 cm$^2$ for both the VSD and Potts shunt, since this appears to the be the smallest possible shunt size that corresponds to the largest decrease in pulmonary artery pressure in either case; refer to Figure \ref{fig:pasa}. Figure \ref{fig:exercise1} shows systemic flow, oxygen saturation, and oxygen delivery as functions of the oxygen consumption, the independent variable in our exercise model. Note that larger consumption indicates a higher level of exercise. Systemic venous oxygen saturation will be our measure of exercise tolerance. It is an important variable to consider during exercise because it indicates the amount of oxygen left after consumption by the systemic tissues, including the exercising muscles. If the computed value for the systemic venous oxygen saturation is negative, then the assumed level of exercise in our model corresponding to that value of oxygen consumption is not possible. 

The left panel of Figure \ref{fig:exercise1} shows systemic flow for each of these four cases. In all cases, systemic flow increases as exercise level increases, as expected. The rate of increase is smaller for the RPH models, both pre- and post-intervention.  We note that the rate of increase in systemic flow for the Pott shunts appears to be slightly larger than that observed for the VSD. The right and bottom panels of Figure \ref{fig:exercise1} show oxygen saturation and oxygen delivery, respectively. The dashed-dotted lines correspond to variables in the systemic arteries and the solid lines correspond to variables in the systemic veins. In all cases, oxygen saturation and delivery decrease as exercise level increases. More rapid decreases in both of these variables are seen for the pre-intervention and post-intervention VSD and Potts shunt models. This trend appears consistent with the notion that the body is less tolerant to exercise in a diseased state. 

More surprisingly, saturation and delivery curves for the VSD and Potts shunt lie beneath the pre-intervention curves, indicating slightly lower exercise tolerances for the post-intervention models compared to the pre-intervention model. This finding is inconsistent with anecdotal evidence that these interventions increase exercise tolerance in RPH patients. In this light, we hypothesize the following mechanism for an increase in exercise tolerance post intervention. First, the intervention such as a Potts shunt or VSD off-loads the right heart and pulmonary vasculature by decreasing the pulmonary artery pressure. Then, {\em remodeling} in the pulmonary artery occurs, leading to a decrease in the pulmonary resistance. Such a change in pulmonary resistance has been observed in Potts shunt patients \cite{Lancaster21}. We test this hypothesis in our models by decreasing the pulmonary resistance. These experiments are done at rest and at moderate exercise corresponding to oxygen consumption values of 16.8 mmol/min and 33.44 mmol/min respectively.

\begin{figure}[h!]
    \centering
    \includegraphics[width=0.48\textwidth]{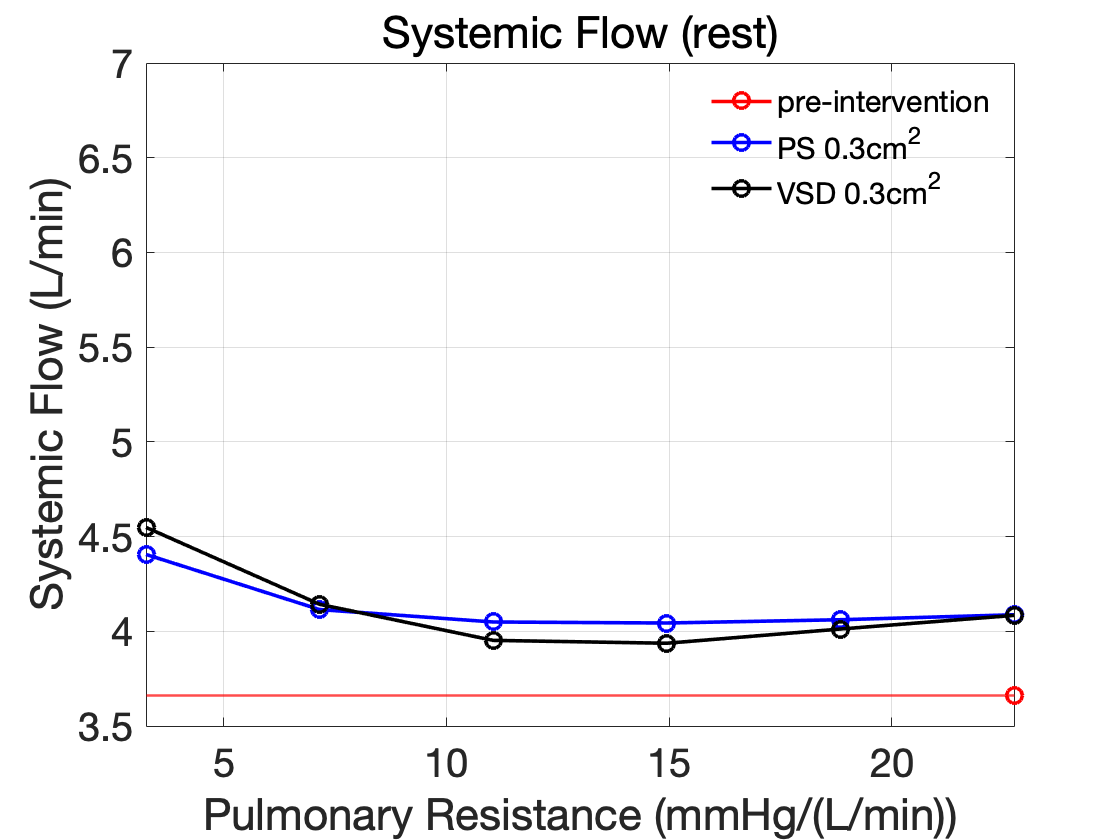}
    \includegraphics[width=0.48\textwidth]{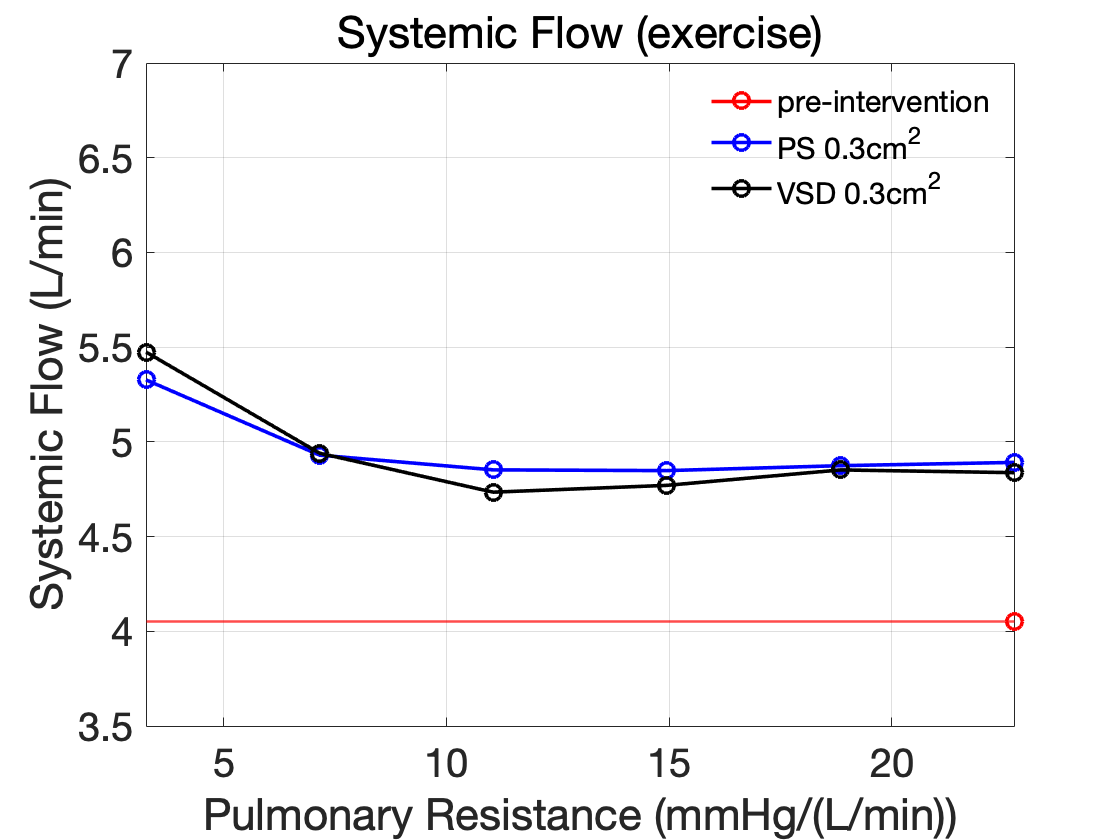}
     \caption{Systemic flow at rest and at moderate exercise. Post-intervention values are shown as the pulmonary resistance decreases in order to simulate post-intervention pulmonary vascular remodeling. The pre-intervention value is shown as a baseline for comparison. The values used for oxygen consumption at rest and at moderate exercise are $M$ = 16.8 mmol/min and $M$ = 33.44 mmol/min respectively.}
    \label{fig:exercise_sf}
\end{figure}

\begin{figure}[h!]
    \centering
    \includegraphics[width=0.48\textwidth]{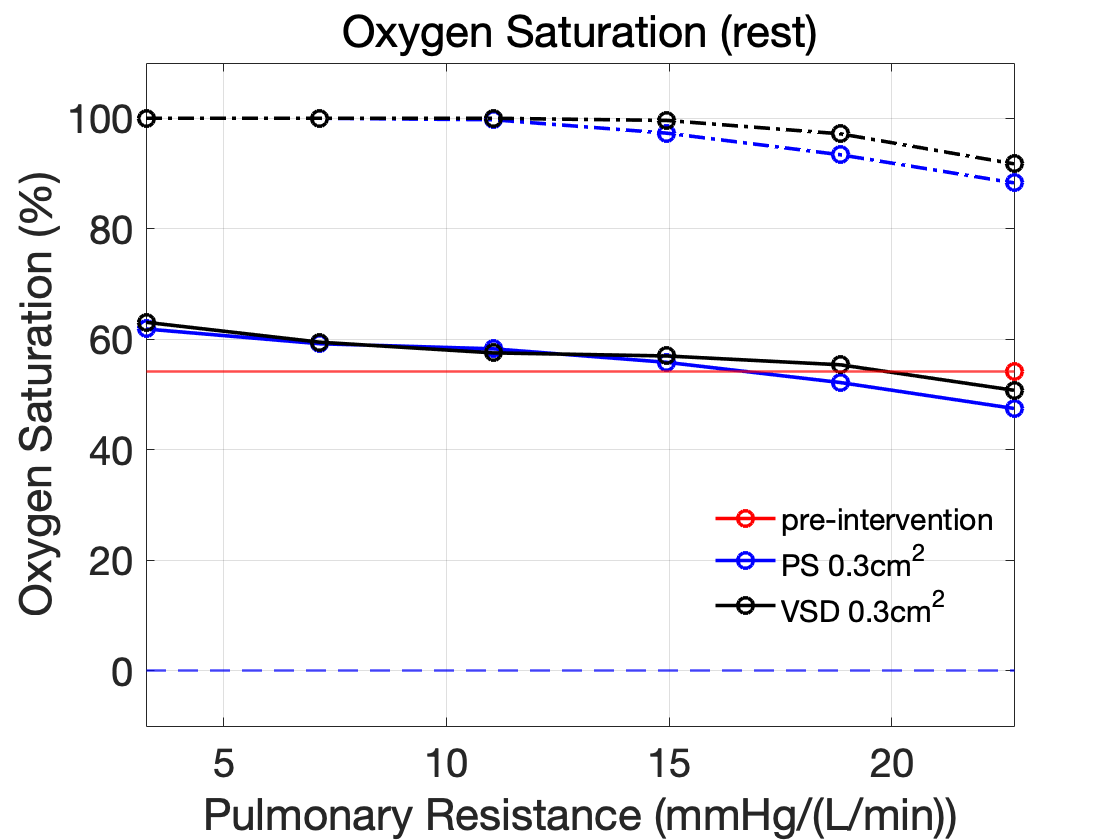}
    \includegraphics[width=0.48\textwidth]{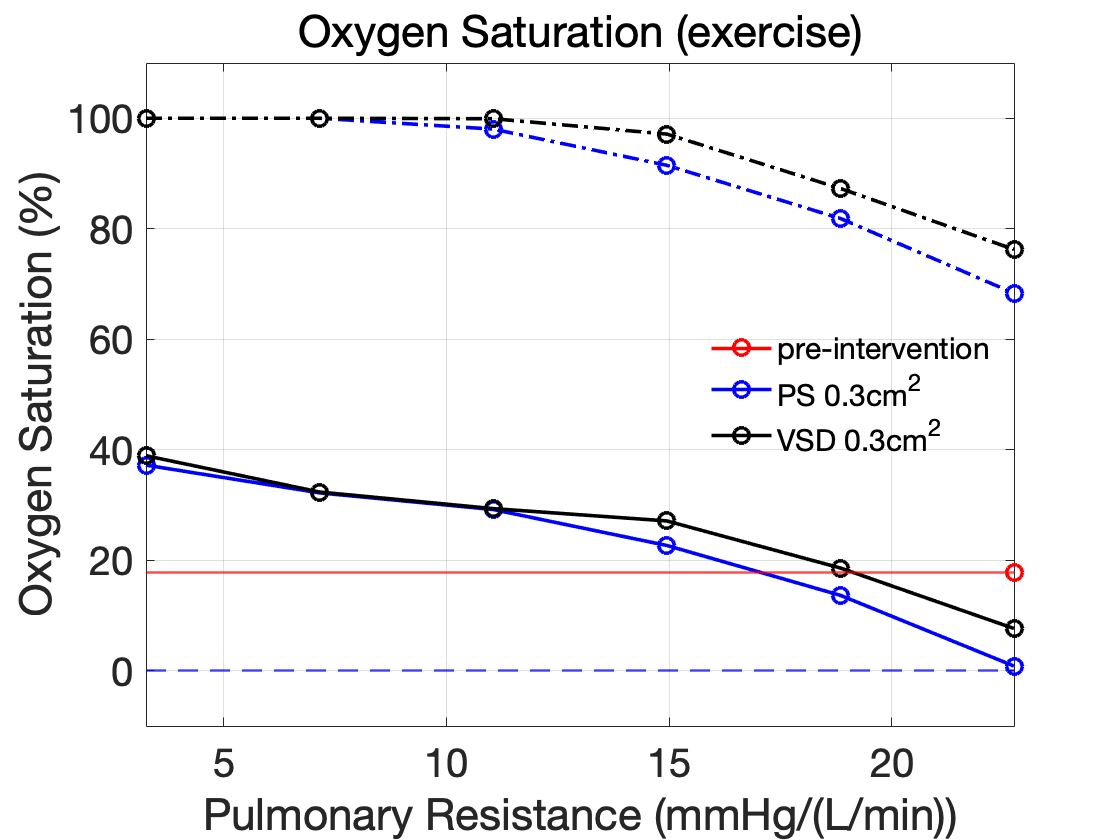}
     \caption{Oxygen Saturation at rest and at moderate exercise. Post-intervention values are shown as the pulmonary resistance decreases in order to simulate post-intervention pulmonary vascular remodeling. The pre-intervention value is shown as a baseline for comparison. The values used for oxygen consumption at rest and at moderate exercise are $M$ = 16.8 mmol/min and $M$ = 33.44 mmol/min respectively.}
    \label{fig:exercise_os}
\end{figure}

\begin{figure}[h!]
    \centering
    \includegraphics[width=0.48\textwidth]{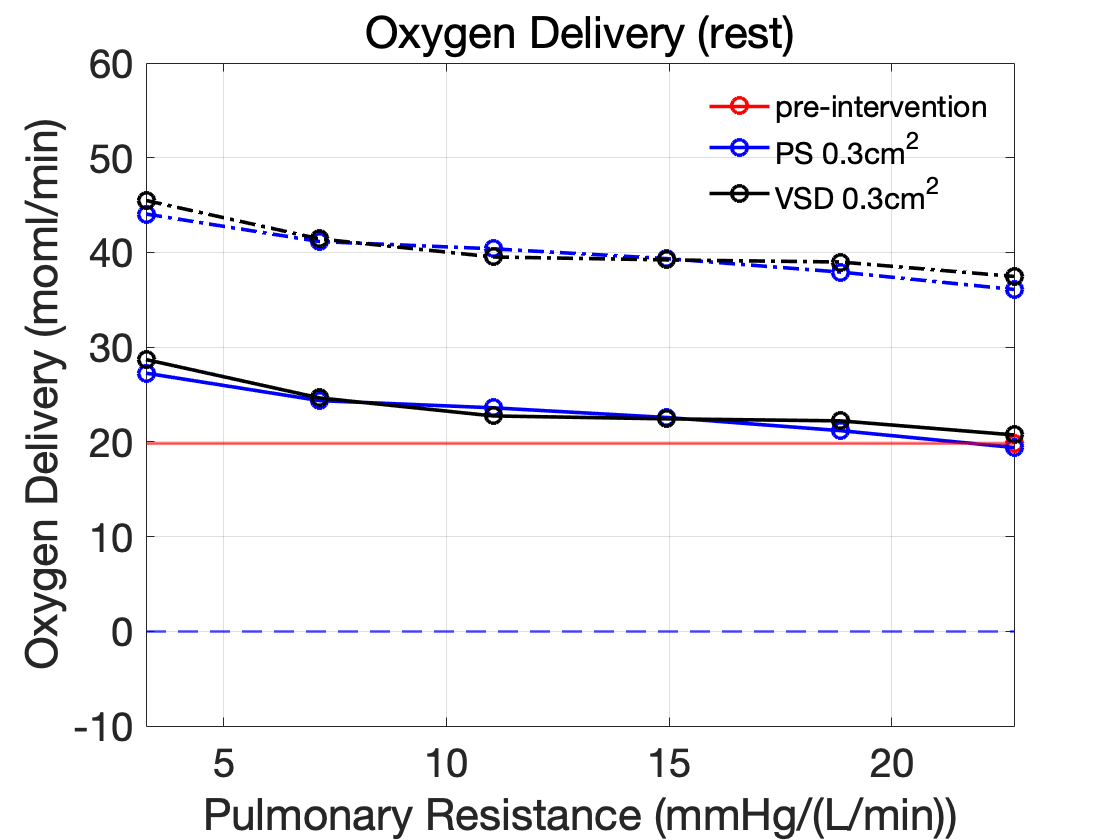}
    \includegraphics[width=0.48\textwidth]{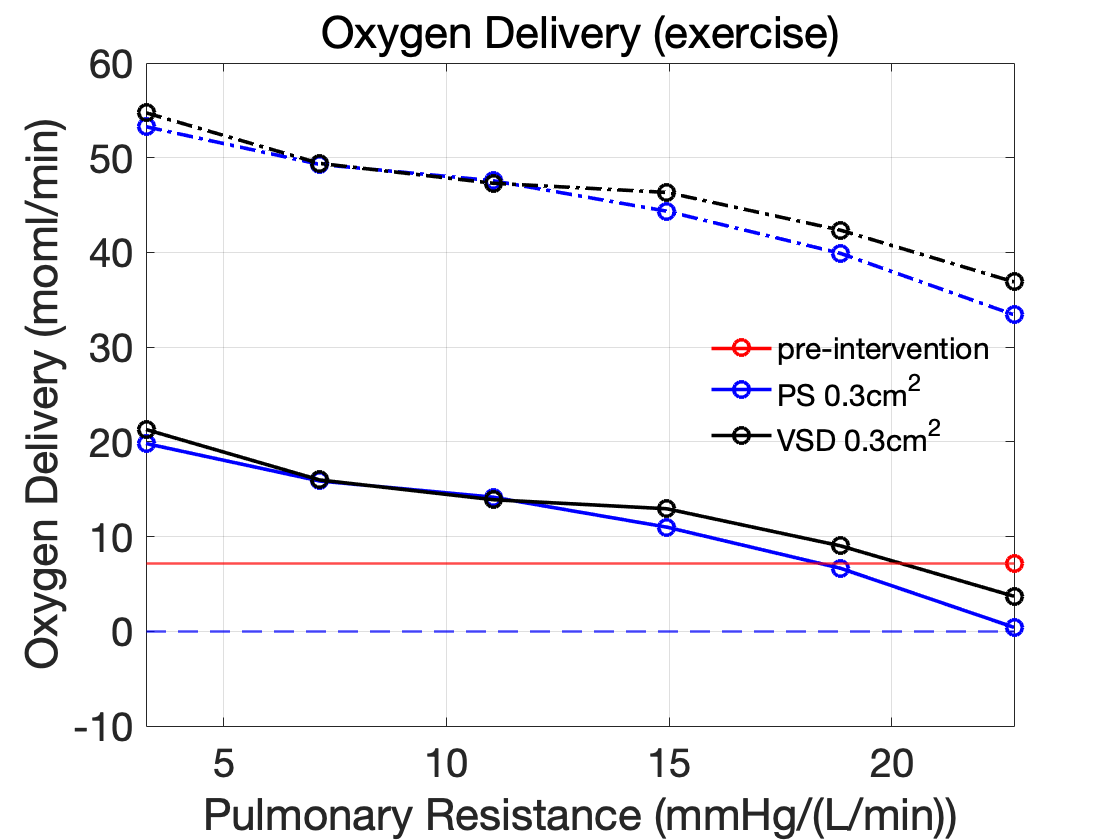}
     \caption{Oxygen Delivery at rest and at moderate exercise. Post-intervention values are shown as the pulmonary resistance decreases in order to simulate post-intervention pulmonary vascular remodeling. The pre-intervention value is shown as a baseline for comparison. The values used for oxygen consumption at rest and at moderate exercise are $M$ = 16.8 mmol/min and $M$ = 33.44 mmol/min respectively.}
    \label{fig:exercise_od}
\end{figure}

Figures \ref{fig:exercise_sf}, \ref{fig:exercise_os}, and \ref{fig:exercise_od} show the effects of decreasing the pulmonary resistance, corresponding to favorable pulmonary vascular remodeling, on systemic flow, oxygen saturation, and oxygen delivery respectively. The model at rest is shown in the left panel and the model corresponding to moderate exercise is shown in the right panel. For each of these three variables, only one data point is shown for the pre-intervention model which has a fixed pulmonary resistance of 22.75 mmHg/(L/min). This point is used as a baseline to assess the performance of the post-intervention models as the pulmonary resistance decreases. For decreasing pulmonary resistance, we see an increase in systemic flow, oxygen saturation, and oxygen delivery. Systemic venous oxygen delivery and saturation for the post-intervention models exceeds the pre-intervention model values for pulmonary resistances less than approximately 15-20 mmHg/(L/min). These results confirm that the VSD or Potts shunt intervention, combined with pulmonary vascular remodeling in the form of decreased pulmonary resistance, lead to an increase in exercise tolerance over the pre-intervention state. We also remark that Figure 7 reveals that the systemic arterial oxygen saturation reaches 100\% for small enough pulmonary resistance, i.e. for substantial enough pulmonary vascular remodeling. Maximum saturation in the systemic arteries indicates the shunt flow has reversed and is now left-to-right. An interesting question is whether the shunt should be closed at this stage, since it apparently serves no therapeutic purpose.


\subsection{Shunt flow waveforms}

\begin{figure}[h!]
    \centering
    \includegraphics[width=0.48\textwidth]{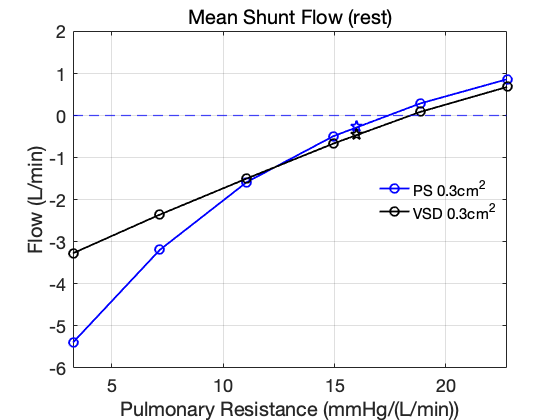}
    \includegraphics[width=0.48\textwidth]{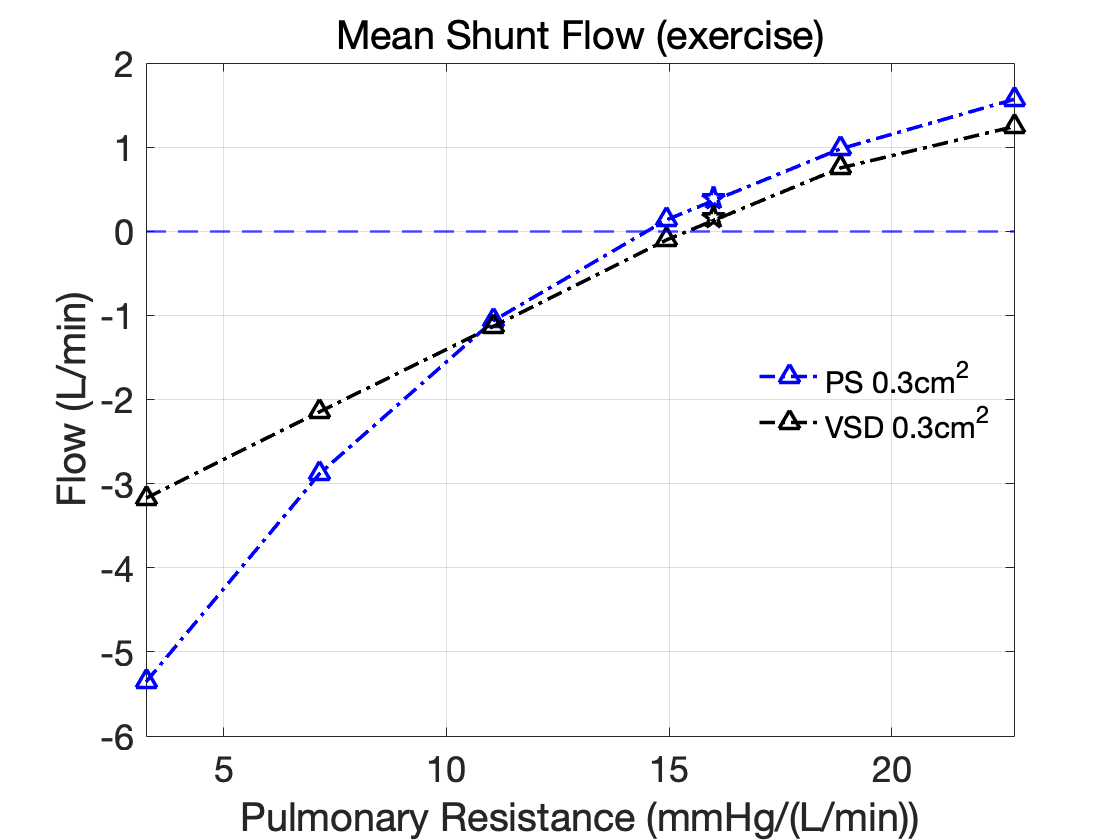}
    \caption{Mean shunt flow for the VSD and Potts shunt, at rest ($M$ = 16.8 mmol/min) and at moderate exercise ($M$ = 33.44 mmol/min), as the pulmonary resistance is varied. The shunt size in both cases is 0.3 cm$^2$. The star-shaped marker corresponds to a pulmonary resistance value of $R_{P}$ = 16  mmHg/(L/min), which is used for the waveforms in Figure \ref{fig:waveform}.}
    \label{fig:meanshunt}
\end{figure}

\begin{figure}[h!]
    \centering
    \includegraphics[width=0.48\textwidth]{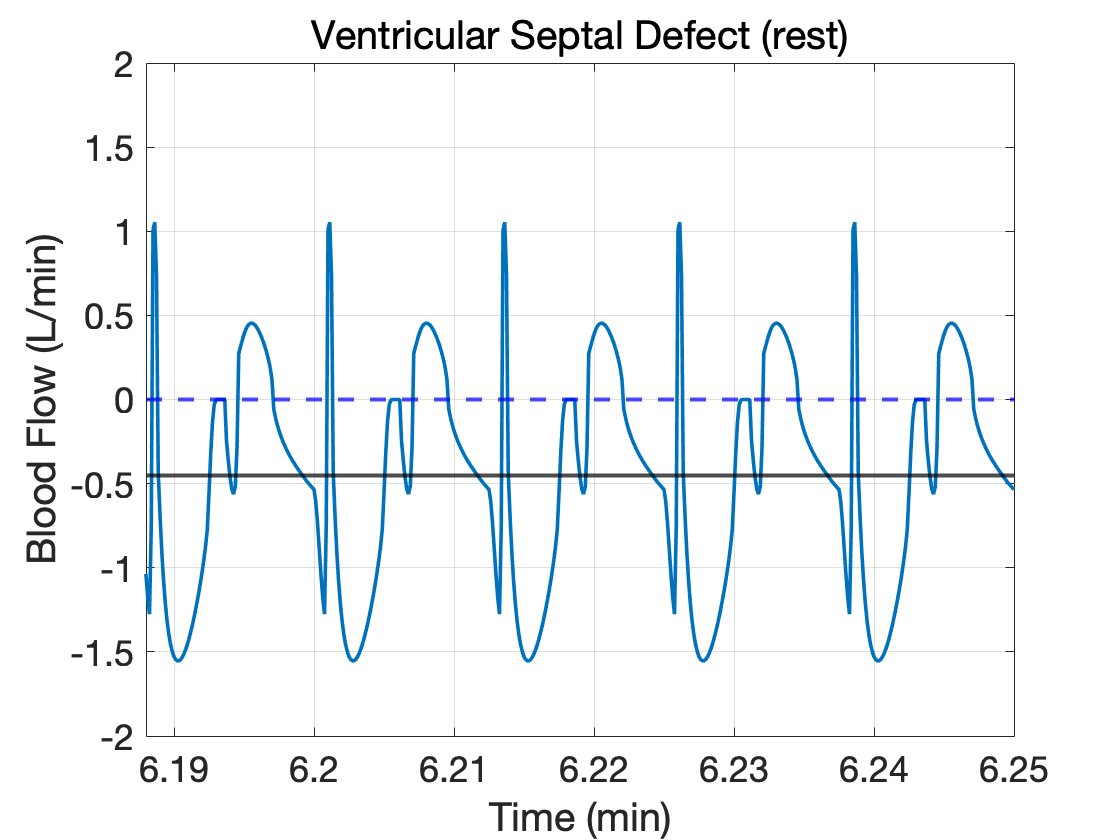}
    \includegraphics[width=0.48\textwidth]{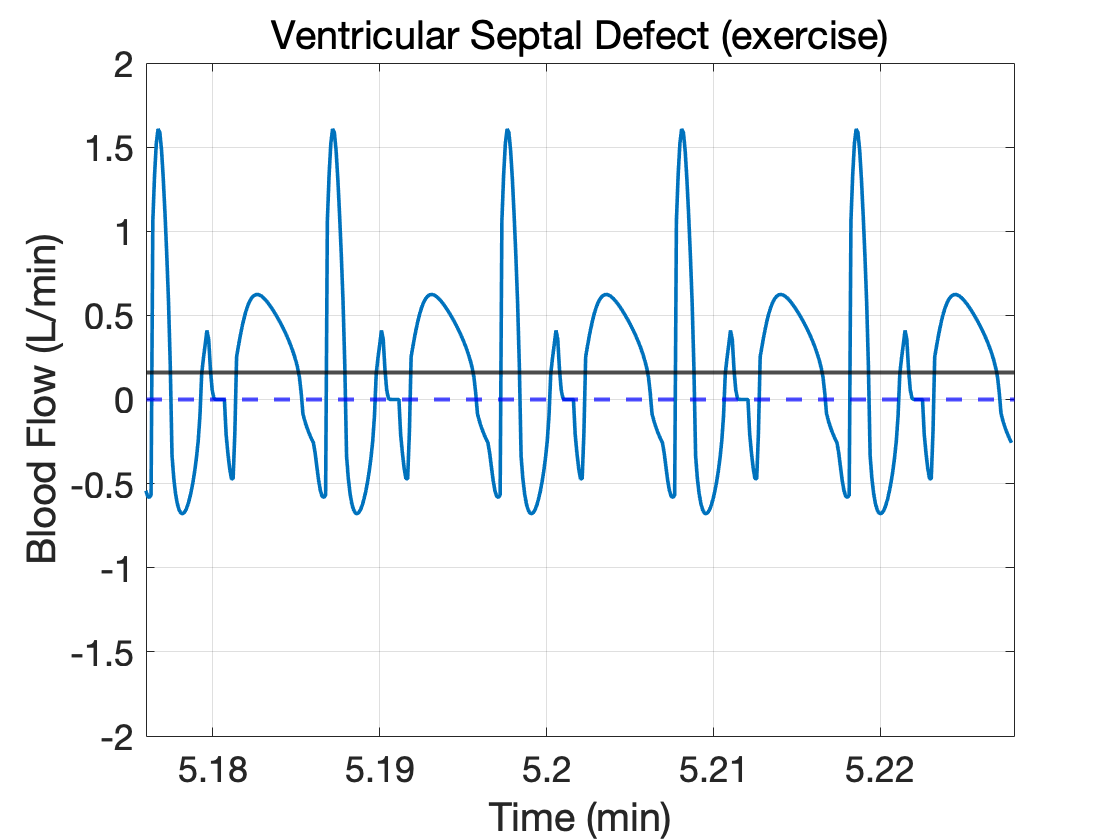}
    \includegraphics[width=0.48\textwidth]{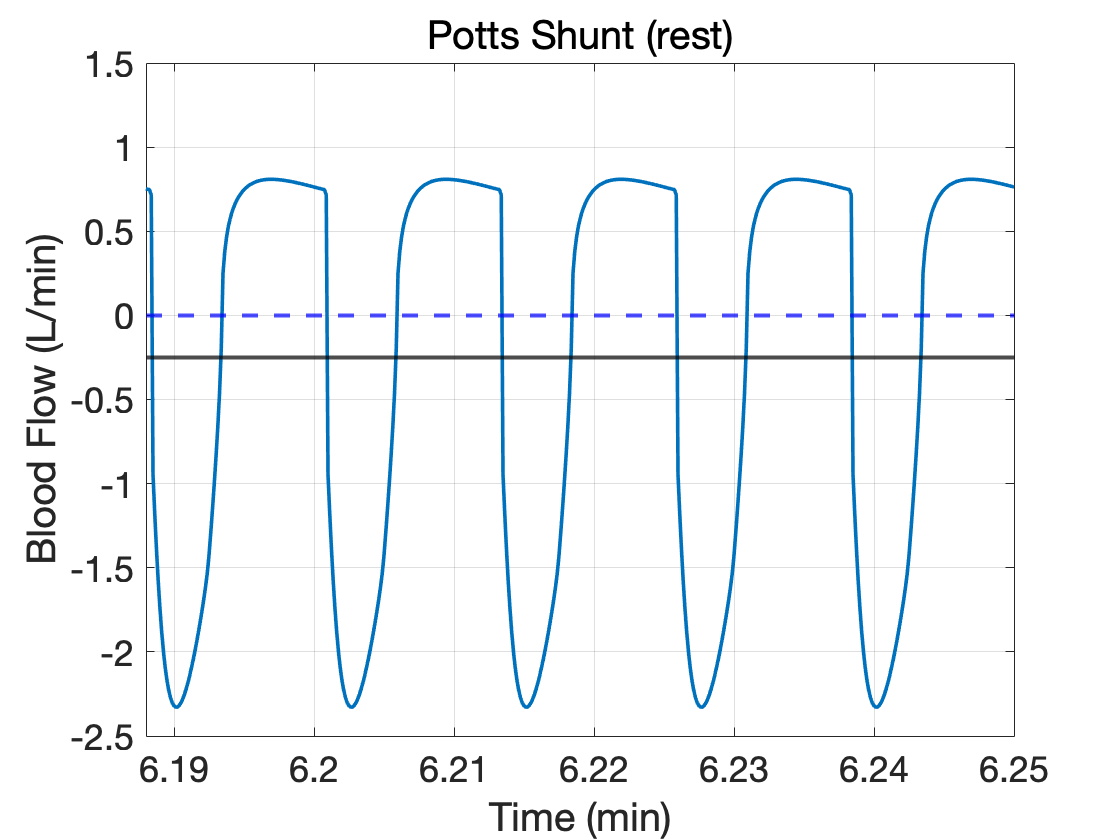}
    \includegraphics[width=0.48\textwidth]{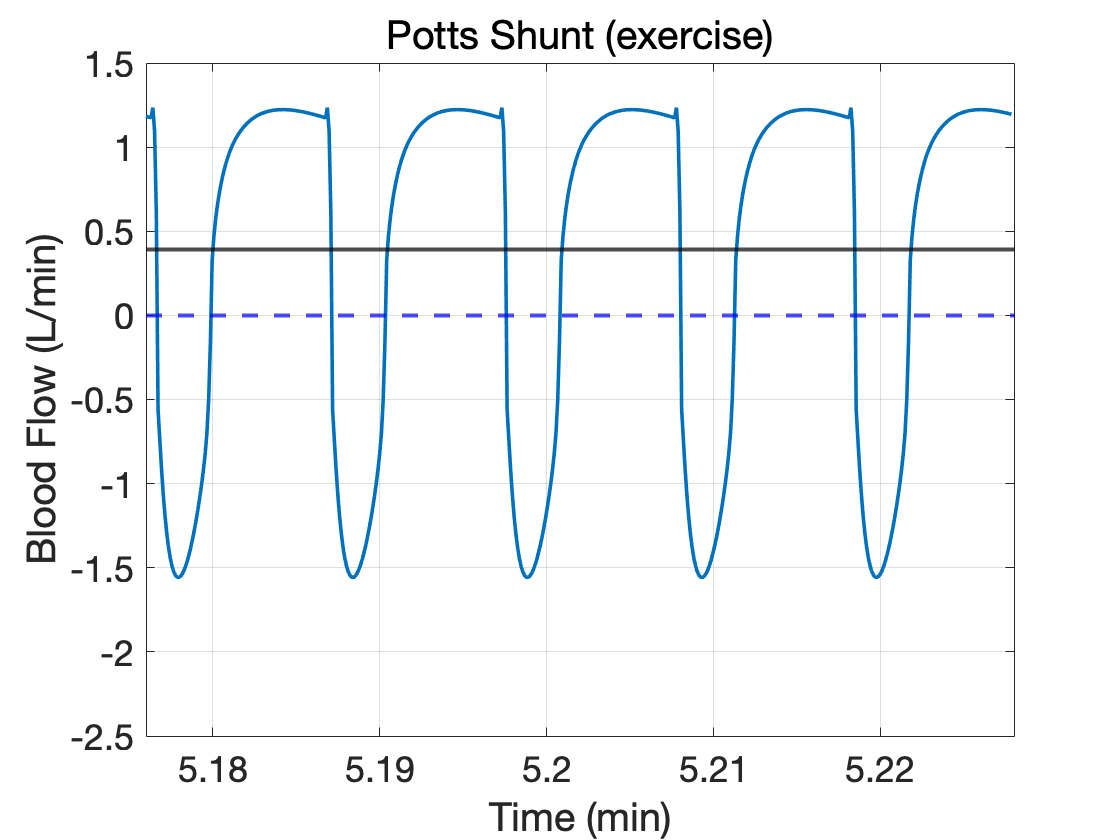}
    \includegraphics[width=0.48\textwidth]{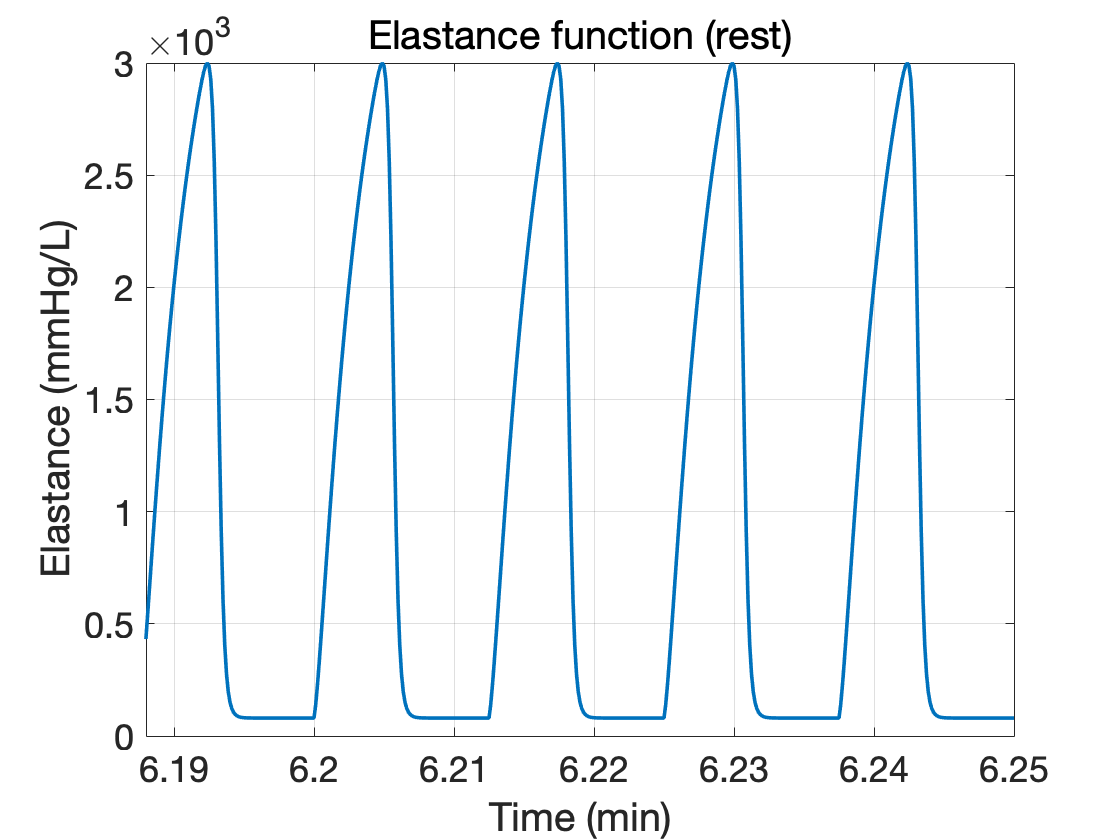}
    \includegraphics[width=0.48\textwidth]{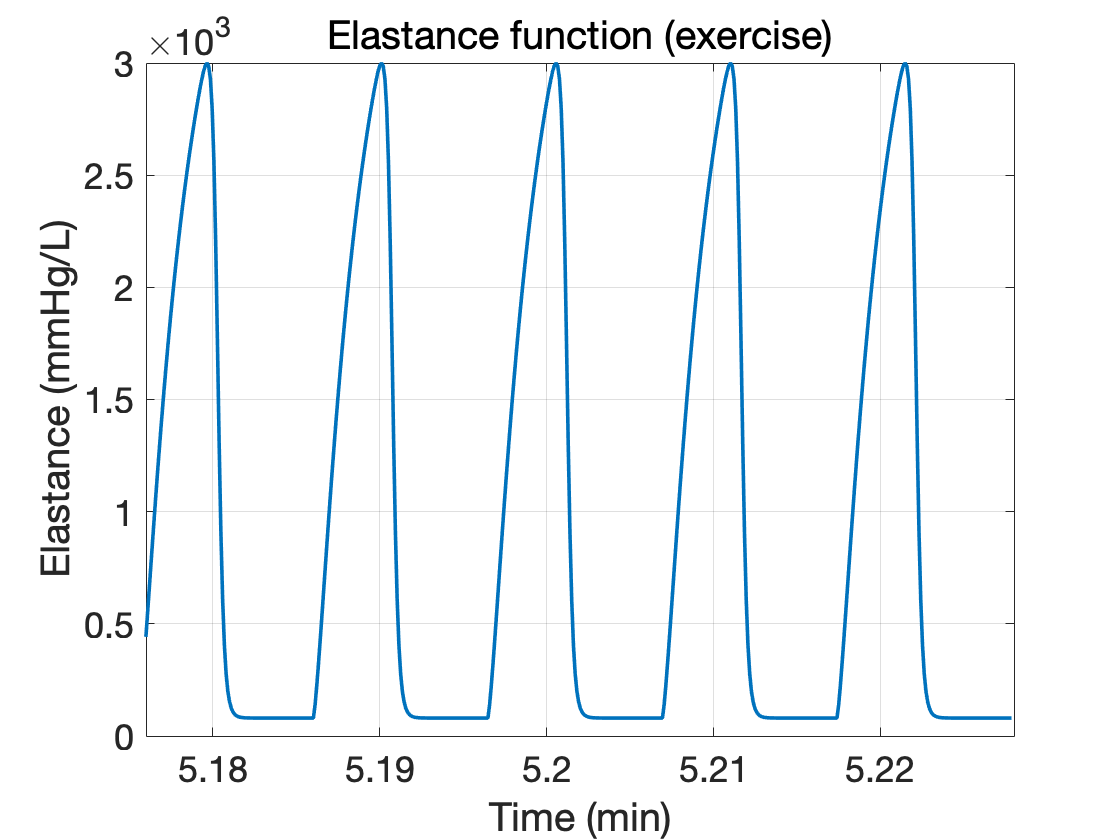}
     \caption{Shunt flow waveforms at rest ($M$ = 16.8 mmol/min) on the left and at moderate exercise ($M$ = 33.44 mmol/min) on the right. The bottom two figures show the elastance function of the ventricle for reference to the cardiac cycle. Note the different time scales; the heart rate is 80 beats/min for the resting case (left) and 95.6 beats/min for the exercise case (right). The shunt size in both cases is 0.3 cm$^2$ and the pulmonary resistance is $R_{P}$ = 16  mmHg/(L/min). The black solid line represents the mean flow over a cardiac cycle of each shunt flow waveform.}
    \label{fig:waveform}
\end{figure}

In this section we examine the flow waveforms through the VSD and Potts shunt. Shunt sizes of 0.3 cm$^2$ are considered for both cases. Figure \ref{fig:meanshunt} depicts the mean shunt flow, as the pulmonary resistance varies, for both interventions. Both rest and exercise are considered. For each shunt, the mean flow switches from right-to-left to left-to-right as the pulmonary resistance decreases. For small enough pulmonary resistances corresponding to favorable remodeling, the mean shunt flow waveforms are left-to-right in both rest and exercise. 
As mentioned above, it might be appropriate in this scenario to close the shunt, since it does not serve any therapeutic benefit. To examine features of the shunt flow waveform, we fix the pulmonary resistance to 16 mmHg/(L/min), a value close to the transition in the mean shunt flow direction (refer to Figure \ref{fig:meanshunt}). Note that at this value for the pulmonary resistance, the oxgyen saturation and delivery levels for the VSD and Potts shunt surpass the pre-intervention values; refer to Figures \ref{fig:exercise_os} and \ref{fig:exercise_od}. In Figure \ref{fig:waveform}, VSD and Potts shunt flow waveforms are shown, with the rest cases on the left and the exercise cases on the right. The black solid line in Figure \ref{fig:waveform} indicates the mean flow value for the waveform over a cardiac cycle. The elastance function for the ventricles is shown in the bottom panel for reference to the cardiac cycle. Note that shunt flows in all cases are complex and bidirectional. However, there is more right-to-left flow during exercise for both the VSD and Potts shunt. For this particular value of the pulmonary resistance, the mean shunt flow is left-to-right at rest and right-to-left during exercise. This finding is consistent with evidence from post-intervention RPH patients who have normal arterial saturations at rest and lower-than-normal saturations during exercise.


As seen in the shunt flow waveforms, bidirectional flow plays an important role in this study. It is an important feature of our methodology that it evaluates the shunt flow as a function of time, and not merely the mean flow. This is because bidirectional flow can exchange oxygen between two compartments even when there is no mean shunt flow at all, and this can have a substantial impact on oxygen transport. Indeed, in the congenital heart disease called transposition of the great arteries, the pulmonary and systemic circulations form parallel loops, and there cannot be any mean flow from one to the other. Survival of the patient after birth is then completely dependent on the existence of a bidirectional shunt \cite{Tu89}.

\section{Limitations}
\label{sec:limitations}
The specific results reported above may certainly depend upon the specific parameters chosen.  In order to apply the methodology of this paper with confidence to any particular patient, it will be necessary to identify the relevant cardiovascular parameters of that patient. Making the model patient-specific is also a way to test the validity of the model, since the pre-operative state of a patient can be used to identify patient-specific parameters, and then the model can be used to predict what the immediately post-operative state of the patient will be.  Comparison with the actual post-operative state will then be a strong test of the model. Future work will therefore be directed toward the development of a methodology for identifying the model parameters that correspond best to the state of a particular patient, so that the model can be made useful in clinical practice. Another limitation of our models is the representation of the systemic arteries as a single compartment. In the case of a VSD, this shunt results in blood mixing in the ventricles, and consequently, desaturated blood is delivered to the brain, coronaries arteries, and lower body. In contrast, the Potts shunt delivers desaturated blood only to the lower body, while fully saturated blood is delivered to the brain and heart. This is by virtue of the fact that mixing occurs downstream from the carotid arteries. This important detail could be studied by constructing a more complex model with additional compartments or by post-processing data from our current model with knowledge of upper and lower systemic arterial compartment blood volumes.

\section{Conclusions}
\label{sec:conc}
In this paper, we have presented a methodology that can be used to study surgical interventions that are designed to alleviate the detrimental effects of refractory pulmonary hypertension. We have illustrated the use of this methodology by comparing three such interventions, all of which are designed to allow some blood flow to bypass the lungs: an atrial septal defect, a ventricular septal defect, and a Potts shunt. For each intervention, we have simulated a range of defect sizes from 0 to 1 cm$^2$. Our results are that the ASD is ineffective at lowering blood pressure in the pulmonary artery, but that the VSD and Potts shunt are both effective, with a greater effect being produced by the Potts shunt.  These results are consistent with the fact that an ASD is volume-unloading while a VSD or Potts shunt is pressure-unloading. Both the VSD and Potts shunt lower the systemic arterial oxygen saturation in our study, but this is partially compensated by an increase in systemic flow, so that oxygen delivery to the systemic tissues is lowered to a lesser degree than the systemic arterial oxygen saturation. The increase in systemic flow is slightly greater for the VSD than for the Potts shunt, with the result that oxygen delivery is slightly increased only in the VSD case. Oxygen delivery is reduced in the case of the Potts shunt due to the substantial oxgyen saturation reduction for Potts shunt compared to VSD. With respect to exercise, both the VSD and Potts shunt saw a reduction in exercise tolerance compared to the pre-intervention case. We found that post-intervention pulmonary vascular remodeling, leading to a drop in pulmonary resistance, explained the anecdotal increase in exercise tolerance that is seen in some RPH patients. When judged by reduction of pulmonary arterial pressure alone, the Potts shunt appears to perform better. When considering oxygen delivery and exercise tolerance, the VSD appears to be the best choice. As mentioned in Section \ref{sec:limitations}, it is important to keep in mind that the Potts shunt has the advantage of delivering fully saturated blood to the brain. The above effects are quantified in our study as functions of the size of the defect in each case, and this kind of information could be useful to a clinician who needs to decide how large a defect to create. 

\section{Acknowledgements}
Charles Puelz was supported in part by the Research Training Group in Modeling and Simulation funded by the National Science Foundation via grant RTG/DMS-1646339.

\bibliographystyle{plain}
\bibliography{references}


\end{document}